\documentclass[a4paper,11pt]{article}
\usepackage[latin1]{inputenc}
\usepackage{amssymb}

\title{ {\bf On  the length of the Wadge  hierarchy   
of \\ $\om$-context free languages}
\footnote{A short version of this paper  appeared in the Proceedings of 
the International Workshop on  Logic and Complexity in Computer Science held  in honour of 
Anatol Slissenko for his $60^{th}$ birthday, Cr\'eteil, France,  2001, \cite{fiwa1}.} }
\author{Olivier Finkel\\
{\it Equipe de Logique Math\'ematique},
\\ U.F.R. de Math\'ematiques, Universit\'e Paris 7 
\\ {\it  2 Place Jussieu 75251 Paris
 cedex 05, France.}
\\{\small E Mail: finkel@logique.jussieu.fr }}
\date{}

\begin{document}

\newtheorem{The}{Theorem}[section]
\newtheorem{Pro}[The]{Proposition}
\newtheorem{Deff}[The]{Definition}
\newtheorem{Lem}[The]{Lemma}
\newtheorem{Rem}[The]{Remark}
\newtheorem{Cor}[The]{Corollary}
\newtheorem{Exa}[The]{Example}

\newcommand{\fa}{\forall}
\newcommand{\Ga}{\Gamma}
\newcommand{\Gas}{\Gamma^\star}
\newcommand{\Si}{\Sigma}
\newcommand{\Sis}{\Sigma^\star}
\newcommand{\Sio}{\Sigma^\omega}

\newcommand{\ra}{\rightarrow}
\newcommand{\hs}{\hspace{12mm}

\noi}
\newcommand{\lra}{\leftrightarrow}
\newcommand{\la}{language}
\newcommand{\ite}{\item}
\newcommand{\Lp}{L(\varphi)}
\newcommand{\abs}{\{a, b\}^\star}
\newcommand{\abcs}{\{a, b, c \}^\star}
\newcommand{\ol}{ $\omega$-language}
\newcommand{\orl}{ $\omega$-regular language}
\newcommand{\om}{\omega}
\newcommand{\nl}{\newline}
\newcommand{\noi}{\noindent}
\newcommand{\tla}{\twoheadleftarrow}
\newcommand{\de}{deterministic }
\newcommand{\vp}{ffi}
\newcommand{\proo}{\noi {\bf Proof.} }
\newcommand {\ep}{\hfill $\square$}

\maketitle

\begin{abstract}
\noi  We prove in this paper that the length of the Wadge hierarchy of 
$\om$-context free languages  
is greater than the Cantor ordinal $\varepsilon_\om$, which is 
the $\om^{th}$ fixed point of the ordinal exponentiation of base $\om$.
 We show also that there exist some 
${\bf \Si^0_\om }$-complete $\om$-context free languages, improving previous results 
on $\om$-context free languages and the Borel hierarchy. 

\end{abstract}

\noi {\small {\bf Keywords:}  $\omega$-context free languages; infinitary  
context free languages; topological properties; Borel hierarchy; 
Wadge hierarchy; conciliating Wadge hierarchy.}

\section{Introduction}

In the sixties  B\"uchi studied the \ol s accepted   by finite   automata to 
prove the decidability of the monadic second order theory of one successor
 over the integers. Since then  the so called $\om$-regular languages have been intensively
 studied, see \cite{tho,pp01} for many results and references. 
The extension to \ol s accepted by pushdown automata has also been investigated, firstly by 
Cohen and Gold, Linna, Nivat,  see Staiger's paper \cite{sta} for a survey of this work, 
including acceptance of infinite words by more powerful accepting devices, 
like Turing machines. A way to investigate the complexity of 
\ol s is to consider their topological complexity. 
Mc Naugthon's Theorem implies that  \orl s  are boolean 
combinations of ${\bf \Pi^0_2 }$-sets.  We proved 
that $\om$-context free languages 
(accepted by pushdown automata with a B\"uchi or Muller acceptance 
condition) exhaust the finite 
ranks  of the Borel hierarchy, \cite{fi}, that there exist some 
$\om$-context free languages ($\om$-CFL) which are analytic but non Borel sets, \cite{f00}, and 
that there exist also some $\om$-CFL which are Borel sets of infinite rank, \cite{f01}. 
\nl On the other side the Wadge hierarchy of Borel sets 
is a great refinement of the Borel hierarchy 
and it induces on \orl s  the now called Wagner hierarchy which has been 
 determined by Wagner in an effective way
 \cite{wag}. Its length is the 
ordinal $\om^\om$.  Notice that Wagner originally determined this hierarchy without 
citing links with the Wadge hierarchy. 
The applicability of the Wadge hierarchy to the Wagner hierarchy 
was first established by Selivanov in \cite{se94,se95}.
The  Wadge hierarchy of {\bf \de} $\om$-context free languages 
has been recently 
determined by Duparc: its length is the ordinal  $\om^{(\om^2)}$,  \cite{dupcf}.  
We proved in \cite{fiwa} that the length of the Wadge hierarchy  
 of (non deterministic) $\om$-context free languages  is an ordinal greater 
than or equal to the  first 
fixed point of the ordinal exponentiation of base $\om$, the  Cantor 
ordinal $\varepsilon_0$. 
\nl
 We  improve here this result and show that the length 
of the Wadge hierarchy  
 of $\om$-context free languages  is an ordinal strictly greater than   
the $\om^{th}$ fixed point 
of the ordinal exponentiation of 
base $\om$, the
ordinal $\varepsilon_\om$. 
 In order to get our results, we  use recent results of  Duparc. 
In  \cite{dup,dup95} he gave a normal form of Borel ${\bf \Delta^0_\om }$-sets, 
i.e. an inductive construction  
of a Borel set of every given degree in the Wadge 
hierarchy of ${\bf \Delta^0_\om }$-Borel sets. 
In the course of the proof he studied the conciliating 
hierarchy which is a hierarchy of sets of finite {\bf and } infinite sequences,  
closely connected to the Wadge hierarchy of non self dual sets.
 On the other hand the infinitary languages, i.e.  
languages containing finite  
{\bf and } infinite words, accepted by pushdown automata have been studied in 
\cite{beaa,beab} where  Beauquier considered these languages as process behaviours 
which may or may not terminate, as for transition systems studied in \cite{an}.
We  study  the conciliating hierarchy of 
infinitary context free  languages, considering various operations over conciliating sets 
and their  counterpart:  arithmetical operations over   Wadge degrees.   
\nl
 On the other side we show that there exists some 
${\bf \Si^0_\om }$-complete $\om$-context free language, using results of descriptive 
set theory on sets of $\om^2$-words and a coding of $\om^2$-words by $\om$-words. 
\nl
The paper is organized as follows. 
In section 2 we recall some above definitions and results about $\om$-languages accepted by 
B\"uchi or Muller pushdown automata.   
In section 3 Borel and Wadge hierarchies are introduced. 
In section 4  we show that the class of infinitary context free 
languages is closed under various operations and we study  the effect of 
these operations on the  Wadge degrees. 
In section 5 we prove our main  result on the length of the Wadge hierarchy 
of $\om$-context free languages. 
In section 6 we construct some ${\bf \Si^0_\om }$-complete $\om$-context free language.

\section{$\om$-Regular and  $\om$-Context Free  Languages}

 We assume the reader to be familiar with the theory of formal \la s and 
of \orl s, \cite{tho,sta}.
 We shall use usual notations of formal language theory. 
  When $\Si$ is a finite alphabet, a {\it non-empty finite word} over $\Si$ is any 
sequence $x=a_1\ldots a_k$ , where $a_i\in\Sigma$ 
for $i=1,\ldots ,k$ ,and  $k$ is an integer $\geq 1$. The {\it length}
 of $x$ is $k$, denoted by $|x|$ .
 The {\it empty word} has no letter and is denoted by $\lambda$; its length is $0$. 
 For $x=a_1\ldots a_k$, we write $x(i)=a_i$  
and $x[i]=x(1)\ldots x(i)$ for $i\leq k$ and $x[0]=\lambda$.
 $\Sis$  is the set of finite words (including the empty word) over $\Sigma$.
 \nl   The first infinite ordinal is $\om$.
 An $\om$-{\it word} over $\Si$ is an $\om$ -sequence $a_1 \ldots a_n \ldots$, where 
$a_i \in\Sigma , \fa i\geq 1$.  When $\sigma$ is an $\om$-word over $\Si$, we write
 $\sigma =\sigma(1)\sigma(2)\ldots \sigma(n) \ldots $,  where for all $i$~ $\sigma(i)\in \Si$,
and $\sigma[n]=\sigma(1)\sigma(2)\ldots \sigma(n)$  for all $n\geq 1$ and $\sigma[0]=\lambda$.
\nl The {\it prefix relation} is denoted $\sqsubseteq$: the finite word $u$ is a {\it prefix}
of the finite word $v$ (respectively,  the infinite word $v$), denoted $u\sqsubseteq v$,  
 if and only if there exists a finite word $w$ 
(respectively,  an infinite word $w$), such that $v=u.w$.
 The set of $\om$-words over  the alphabet $\Si$ is denoted by $\Si^\om$.
An  $\om$-{\it language} over an alphabet $\Sigma$ is a subset of  $\Si^\om$.
 \nl  For $V\subseteq \Sis$, the $\om$-{\it power} of $V$ is the \ol:
$$V^\om = \{ \sigma =u_1\ldots u_n \ldots \in \Si^\om \mid 
 \fa i\geq 1 ~ u_i\in V-\{\lambda\} \}$$

\noi For any family $\mathcal{L}$ of  finitary \la s, the $\om$-{\it Kleene closure}
of $\mathcal{L}$, is : 
$$\om-KC(\mathcal{L}) = \{ \cup_{i=1}^n U_i.V_i^\om  \mid  U_i, V_i \in \mathcal{L}, 
 \fa i\in [1, n] \}$$

 \noi For $V\subseteq \Sis$, the complement of $V$ (in $\Sis$) is $\Sis - V$ denoted $V^-$.
  For a subset $A\subseteq \Si^\om$, the complement of $A$ is 
$\Si^\om - A$ denoted $A^-$.
 When we consider subsets of $\Si^{\leq \om}=\Sis \cup \Si^\om$, 
if $A\subseteq \Si^{\leq \om}$ then $A^-=\Si^{\leq \om}-A$, 
(this will be clear from the context so that there will not be any confusion even if 
$A\subseteq \Sis$ or $A\subseteq \Sio$).

\hs Recall that the class $REG_\om$ of  $\om$-{\it regular languages}  is the class of 
\ol s accepted by finite automata with a B\"uchi or Muller acceptance condition. It is 
also the $\om$-Kleene closure of  the class $REG$ of regular finitary languages. 
\nl Similarly the class $CFL_\om$ of  $\om$-{\it context free languages} ($\om$-CFL) 
 is the class 
of \ol s accepted by pushdown automata with a B\"uchi or Muller acceptance condition. It is 
also the $\om$-Kleene closure of the class $CFL$ of context free  finitary languages, 
\cite{cg,sta}.
\nl Let $\Si$ be a finite alphabet. A subset $L$ 
of $\Si^{\leq \om}$ is said to be an {\it infinitary 
context free language} iff there exists a finitary context free language 
$L_1\subseteq \Si^\star$ 
and an $\om$-CFL $L_2\subseteq \Si^\om$ such that $L=L_1\cup L_2$.  The class of 
infinitary context free languages will be denoted  $CFL_{\leq \om}$.

\section{Borel and Wadge Hierarchies}

\noi We assume the reader to be familiar with basic notions of topology which
may be found in \cite{mos,kec,lt,sta,pp01} 
and  with the elementary theory 
of  ordinals, including the operations of multiplication and exponentiation, 
which may be found in \cite{sier}.
 For a finite alphabet $X$, we consider $X^\om$ 
as a topological space with the {\it Cantor topology}.
 The {\it open sets} of $X^\om$ are the sets in the form $W.X^\om$, where $W\subseteq X^\star$.
A set $L\subseteq X^\om$ is a {\it closed set} iff its complement $X^\om - L$ is an open set.
 Define now the  the {\it Borel Hierarchy} on $X^\om$:

\begin{Deff}
For a non-null countable ordinal $\alpha$, the classes ${\bf \Si^0_\alpha }$
 and ${\bf \Pi^0_\alpha }$ of the Borel Hierarchy on the topological space $X^\om$ 
are defined as follows:
\nl ${\bf \Si^0_1 }$ is the class of open subsets of $X^\om$.
\nl ${\bf \Pi^0_1 }$ is the class of closed subsets of $X^\om$.
\nl and for any countable ordinal $\alpha \geq 2$: 
\nl ${\bf \Si^0_\alpha }$ is the class of countable unions of subsets of $X^\om$ in 
$\cup_{\gamma <\alpha}{\bf \Pi^0_\gamma }$.
 \nl ${\bf \Pi^0_\alpha }$ is the class of countable intersections of subsets of $X^\om$ in 
$\cup_{\gamma <\alpha}{\bf \Si^0_\gamma }$.
\end{Deff}

\noi  Notice that the above  definition of Borel classes ${\bf \Si^0_\alpha }$ 
and ${\bf \Pi^0_\alpha }$, for a {\it limit ordinal} $\alpha$, is the usual one in 
descriptive set theory, as given in the textbooks \cite{mos,kec}. 
\nl In particular, the class ${\bf \Si^0_\om }$ is not the union of the classes ${\bf \Si^0_n}$ 
for  integers $n\geq 1$ but it strictly contains 
 $\cup_{n\geq 1} {\bf \Si^0_n}= \cup_{n\geq 1} {\bf \Pi^0_n}$ consisting 
of Borel sets of finite rank. Moreover classes ${\bf \Si^0_\om }$ and ${\bf \Pi^0_\om }$ are 
distinct and are incomparable for the inclusion relation.  
\nl We shall say that a subset of $X^\om$ is a {\it Borel set of rank} $\alpha$, for 
a countable ordinal $\alpha$,  iff 
it is in ${\bf \Si^0_{\alpha}}\cup {\bf \Pi^0_{\alpha}}$ but not in 
$\bigcup_{\gamma <\alpha}({\bf \Si^0_\gamma }\cup {\bf \Pi^0_\gamma})$. 
\nl In particular a Borel set has Borel rank $\omega$ iff it is in 
$({\bf \Si^0_\om } \cup {\bf \Pi^0_\om })$ but is not a Borel set of finite rank.      

\hs Introduce now the {\it Wadge Hierarchy} which is in fact a huge refinement of 
the Borel hierarchy:

\begin{Deff}[Wadge \cite{wad}] Let $X$, $Y$ be two finite alphabets. 
For $E\subseteq X^\om$ and $F\subseteq Y^\om$, $E$ is said to be Wadge reducible to $F$
($E\leq _W F)$ iff there exists a continuous function $f: X^\om \ra Y^\om$, such that
$E=f^{-1}(F)$.
\nl $E$ and $F$ are Wadge equivalent iff $E\leq _W F$ and $F\leq _W E$. 
This will be denoted by $E\equiv_W F$. And we shall say that 
$E<_W F$ iff $E\leq _W F$ but not $F\leq _W E$.
\nl  A set $E\subseteq X^\om$ is said to be self dual iff  $E\equiv_W E^-$, and otherwise 
it is said to be non self dual.
\end{Deff}

\noi
 The relation $\leq _W $  is reflexive and transitive,
 and $\equiv_W $ is an equivalence relation.
\nl The {\it equivalence classes} of $\equiv_W $ are called {\it Wadge degrees}. 
\nl $WH$ is the class of Borel subsets of a set  $X^\om$, where  $X$ is a finite set,
 equipped with $\leq _W $ and with $\equiv_W $.
\nl 
Remark that in the above definition, we consider that a subset $E\subseteq  X^\om$ is given
together with the alphabet $X$.
\nl We can now define the {\it Wadge class} of a set $F$:

\begin{Deff}
Let $F$ be a subset of $X^\om$. The Wadge class of $F$ is $[F]$ defined by:
$[F]= \{ E \mid  E\subseteq Y^\om$ for a finite alphabet $Y$ and $E\leq _W F \}$. 
\end{Deff}

\noi Recall that each {\it Borel class} ${\bf \Si^0_\alpha}$ and ${\bf \Pi^0_\alpha}$ 
is a {\it Wadge class}.
\nl And that a set $F\subseteq X^\om$ is a ${\bf \Si^0_\alpha}$
 (respectively ${\bf \Pi^0_\alpha}$)-{\it complete set} iff for any set 
$E\subseteq Y^\om$, $E$ is in 
${\bf \Si^0_\alpha}$ (respectively ${\bf \Pi^0_\alpha}$) iff $E\leq _W F $.

\begin{The} [Wadge]
Up to the complement and $\equiv _W$, the class of Borel subsets of $X^\om$,
 for  a finite alphabet $X$, is a well ordered hierarchy.
 There is an ordinal $|WH|$, called the length of the hierarchy, and a map
$d_W^0$ from $WH$ onto $|WH|-\{0\}$, such that for all $A, B\in WH$:
\nl $d_W^0 A < d_W^0 B \lra A<_W B $  and 
\nl $d_W^0 A = d_W^0 B \lra [ A\equiv_W B $ or $A\equiv_W B^-]$.
\end{The}

\noi 
 The Wadge hierarchy of Borel sets of {\bf finite rank }
has  length $^1\varepsilon_0$ where $^1\varepsilon_0$
 is the limit of the ordinals $\alpha_n$ defined by $\alpha_1=\om_1$ and 
$\alpha_{n+1}=\om_1^{\alpha_n}$ for $n$ a non negative integer, $\om_1$
 being the first non countable ordinal. Then $^1\varepsilon_0$ is the first fixed 
point of the ordinal exponentiation of base $\om_1$. The length of the Wadge hierarchy 
of Borel sets in ${\bf \Delta^0_\om}= {\bf \Si^0_\om}\cap {\bf \Pi^0_\om}$ 
  is the $\om_1^{th}$ fixed point 
of the ordinal exponentiation of base $\om_1$, which is a much larger ordinal. The length 
of the whole Wadge hierarchy of Borel sets is a huge ordinal, with regard 
to the $\om_1^{th}$ fixed point 
of the ordinal exponentiation of base $\om_1$. It is described in \cite{wad,dup} 
by the use of the Veblen functions. 
\nl  There is an effective version of the Wadge hierarchy restricted to \orl s:

\begin{The} For $A$ and $B$ some 
$\om$-regular sets,
 one can effectively decide whether $A\leq_W B$ and one can compute $d_W^0(A)$.
\end{The}

\noi The hierarchy obtained on \orl s is now called the Wagner hierarchy and has length 
$\om^\om$.
Wagner \cite{wag} gave an automata structure characterization, based on notion of chain
 and superchain, for an automaton to be in a given class and then he got an 
algorithm to compute the Wadge degree of an \orl . Wilke and Yoo  
proved in \cite{wy} that  one can compute in polynomial time the Wadge degree of an \orl. 
There is an effective extension of the Wagner hierarchy: the Wadge hierarchy of \ol s 
accepted by Muller deterministic one blind (i. e. without zero-test)  counter 
 automata \cite{csl}. 
This  hierarchy has an extension to \de $\om$-context free languages 
as well as to \de Petri net \ol s 
which has length $\om^{(\om^2)}$ \cite{dfr,dupcf,wadpn} but we do not know yet 
whether these extensions are  decidable. 
\nl The Wadge hierarchy of $\om$-languages accepted by {\bf \de} Turing machines 
has been very recently determined by Selivanov: 
its length is the ordinal $(\om_1^{CK})^\om$, where 
$\om_1^{CK}$ is the first non-recursive ordinal,  \cite{se03}.
\nl  The Wadge hierarchy restricted to (non deterministic) 
$\om$-CFL is not effective: we have shown 
in \cite{fi,fiwa,f00} that one can neither decide the Borel rank nor  the Wadge 
degree of a Borel $\om$-CFL. In fact one cannot even decide whether an $\om$-CFL is a 
Borel set. 

\section{Operations on Conciliating  Sets}

\subsection{Conciliating Sets}

\noi We  sometimes consider here subsets of $X^\star \cup X^\om = X^{\leq \om}$, for an 
alphabet $X$, which are called {\it conciliating sets} in \cite{dup,dup95}.

\begin{Deff} Let $X$ be a finite or countably infinite alphabet. A conciliating set 
over the alphabet $X$ is a subset of the set $X^{\leq \om}=X^\star \cup X^\om$ of finite or 
infinite words over $X$. 
\end{Deff}

\begin{Rem}
We shall  only consider in the sequel  conciliating sets defined over a {\bf finite} 
alphabet, except that we shall state Definition \ref{defsup} 
and Proposition \ref{sup} in a more general case. 
\end{Rem}

\noi In order to give a ``normal form" of Borel sets in the Wadge hierarchy, 
J. Duparc studied the 
conciliating (Wadge) hierarchy which is a hierarchy over conciliating sets closely related 
to the Wadge hierarchy. Recall the definition of the {\it conciliating Wadge game}:

\begin{Deff}[\cite{dup}] Let $X_A$, $X_B$ be two finite alphabets. 
 For $A\subseteq X_A^{\leq \om}$ and $B\subseteq X_B^{\leq \om}$,  
the conciliating Wadge game  $C(A, B)$ is a game with perfect information between two players,
player 1 who is in charge of $A$ and player 2 who is in charge of $B$.
\nl Player 1 first writes a letter $a_1\in X_A$, then  
the two players alternatively write letters $a_n$ of $X_A$ for player 1 and $b_n$ of $X_B$
for player 2.
\nl Both players are allowed to skip even indefinitely if they want to. 
\nl Then after $\om$ steps, the player 1 has written a (finite or infinite) word 
$x\in X_A^{\leq \om}$ and the player 2
has written a (finite or infinite) word $y\in X_B^{\leq \om}$.
\nl Player 2 wins the play iff [$x\in A \lra y\in B$], i.e. iff 
[$(x\in A \mbox{ and } y\in B) \mbox{ or } (x\notin A \mbox{ and } y\notin B)$], 
 otherwise player 1 wins. 
\end{Deff}

\noi A {\it strategy} for player 1 is a function 
$\sigma :(X_B\cup \{s\})^\star\ra (X_A \cup \{s\})$.
And a strategy for player 2 is a function 
$f:(X_A\cup \{s\})^+\ra X_B\cup\{ s\}$ \quad ($s$ for "skip").
\nl A strategy 
 $\sigma$ is a {\it winning strategy} (w.s.) for player 1 iff he always wins a play when
 he uses the strategy $\sigma$, i.e. when  he writes at step $n$ the letter 
 $a_n=\sigma (b_1...b_{n-1})$ if $a_n \neq s$ 
and he skips at step $n$ if $s=\sigma (b_1...b_{n-1})$.
\nl A winning strategy for player 2 is defined in a similar manner.
\nl Without loss of generality we can consider that players are allowed 
to write a finite word instead of a single letter at each step of a play \cite{dup}.

\begin{Deff} For $A\subseteq X_A^{\leq \om}$ and $B\subseteq X_B^{\leq \om}$,  
$A\leq_c B$  iff player 2 has a winning strategy in $C(A, B)$. Then 
$A<_c B$  iff $A\leq_c B$ but not conversely 
and $A\equiv_c B$  iff $A\leq_c B \leq_c A $. 
\end{Deff}

\noi 
It turned out that in the conciliating Wadge hierarchy every conciliating 
set is non self dual. The Wadge hierarchy and  the 
conciliating  Wadge hierarchy 
are connected via the following correspondence:
\nl   First define $A^d$ for $A\subseteq \Si^{\leq \om}$ and $d$ a letter not in $\Si$:
 \begin{center}
$A^d = \{ x\in (\Si \cup \{d\})^\om ~\mid  ~x( /d)\in A \}$
\end{center}
where $x( /d)$ is the sequence obtained from $x$ 
when removing every occurrence of the letter $d$.
 Then  for $A\subseteq \Si^{\leq \om}$ such that $A^d$ is a Borel set, (we shall say in that 
case that $A$ is a Borel conciliating set),
$A^d$ is always a non self dual subset of 
$(\Si\cup\{d\})^\om$ and the correspondence $A\ra A^d$ induces an isomorphism between the 
conciliating hierarchy and the Wadge hierarchy of non self dual Borel sets. 
This is due to the 
fact that for two conciliating sets $A$ and $B$,
$$A \leq_c B  \mbox{ iff } A^d \leq_W B^d$$

\noi  Martin's Theorem states that Borel Gale-Stewart games are {\it determined}: 
in such infinite games one of the two players has a winning strategy,  see \cite{kec}. 
This implies the following result. 

\begin{The}\label{det}
 Let $X_A$, $X_B$ be two finite alphabets and 
 $A\subseteq X_A^{\leq \om}$,  $B\subseteq X_B^{\leq \om}$ such that  
$A^d$ and $B^d$ are Borel sets. 
Then the conciliating Wadge game $C(A, B)$  is determined:  
one of the two players has a winning strategy. 
\end{The}

\noi 
From now on we shall first concentrate on non self dual sets as in \cite{dup} 
and we shall use the 
following definition of the Wadge degrees  which is a slight modification of the previous 
one:

\begin{Deff}
\noi
\begin{enumerate}
\ite[(a)]  $d_w(\emptyset)=d_w(\emptyset^-)=1$
\ite[(b)]  $d_w(A)=sup \{d_w(B)+1 ~\mid ~ B {\rm ~non~ self ~dual~ and~}  B<_W A \} $  
\nl (for either $A$ self dual or not, $A>_W \emptyset).$
\end{enumerate}
\end{Deff}

\noi Recall the definition of the {\it conciliating degree} of a conciliating set:

\begin{Deff}
Let $A\subseteq \Si^{\leq \om}$ be a conciliating set over the alphabet $\Si$ such that 
$A^d$ is a Borel set.
The conciliating degree of $A$ is  $d_c(A)=d_w(A^d)$. 
\end{Deff}

\noi We  recall  now some properties of the correspondence $A\ra A^d$ when context free
languages are considered:

\begin{Pro}[\cite{fi}]\label{prod} 
\begin{enumerate}
\ite[a)]  if $A \subseteq \Sis$ is a context free (finitary) 
language, or if $A \subseteq \Si^\om$ is an $\om$-CFL, then $A^d$ is  an $\om$-CFL.
\ite[b)]  If $A$ is the union of a finitary context free  language
and of an $\om$-CFL over the same alphabet $\Si$, then $A^d$ is an $\om$-CFL over the 
alphabet $\Si\cup\{d\}$.
\end{enumerate}
\end{Pro}

\noi  We are going now to 
introduce several operations over conciliating sets: the operation of sum, 
of exponentiation and of iterated exponentiation. And we shall study their counterpart
which are  ordinal 
arithmetical operations over Wadge degrees. 

\subsection{Operation of Sum}

\begin{Deff}[\cite{dup}]
Assume that $X_A\subseteq X_B$ are two finite alphabets,  
  $X_B-X_A$ containing at least two elements, and that
$\{X_+, X_-\}$ is a partition of $X_B-X_A$ in two non empty sets.
 Let $A \subseteq X_A^{\leq \om}$ and $B \subseteq X_B^{\leq \om}$, then 
 $$B + A =_{df} A\cup \{ u.a.\beta  ~\mid  ~ u\in X_A^\star , ~(a\in X_+ ~and ~\beta \in B )~
or ~(a\in X_- ~and ~\beta \in B^- )\}$$
\end{Deff}

\noi This operation is closely related to the {\it ordinal sum}
 as it is stated in the following:

\begin{The}[\cite{dup}]\label{thesum}
Let $X_A\subseteq X_B$, $X_B-X_A$ containing at least two elements,
   $A \subseteq X_A^{\leq \om}$ and $B \subseteq X_B^{\leq \om}$ such 
that $A^d$ and $B^d$ are Borel sets. 
Then $(B+A)^d$ is a Borel set and  
$d_c(B + A )= d_c( B) + d_c( A)$.  
\end{The}

\begin{Rem} As indicated in Remark 5 of \cite{dup},
when $A\subseteq \Si^{\leq \om}$ and $X$ is a finite alphabet, it is easy to build 
$A'\subseteq (\Si\cup X)^{\leq \om}$, such that $(A')^d\equiv_W A^d$. In fact $A'$ can be 
defined as follows:
 for $\sigma \in (\Si\cup X)^{\leq \om}$, let $\sigma \in A' \lra \sigma ' \in A$, where
$\sigma '$ is $\sigma $ except that each letter not in $\Si$ is removed.
Then in the sequel we assume that each alphabet is as enriched as desired, and in particular 
we can always define $B+A$ (or in fact another set $C$ such that $C^d \equiv_W (B+A)^d$). 
\end{Rem}

\noi  Consider now conciliating sets which are union of a finitary  CFL 
 and of an $\om$-CFL. 

\begin{Pro}[\cite{fiwa}]\label{prosum}
Let $X_A\subseteq X_B$ such that 
$\{X_+, X_-\}$ is a partition of $X_B-X_A$ in two non empty sets.
Assume $A\subseteq X_A^{\leq \om}$,  $A, A^- \in CFL_{\leq\om}$, 
 $B\subseteq X_B^{\leq \om}$ and $B, B^- \in CFL_{\leq\om}$. 
 Then $B + A $ and $(B+A)^-$ are in $CFL_{\leq \om}$. 
\end{Pro}

\begin{Deff}
Let $A\subseteq X_A^{\leq \om}$  be a conciliating set over the alphabet $X_A$.
Then $A.n$ is inductively defined by
$A.1=A$ 
and $A.(n+1)=(A.n)+A$,  for each integer $n\geq 1$.
\end{Deff}

\subsection{Operation of Exponentiation}

\noi We are going now to introduce the operation of exponentiation of 
conciliating sets which was firstly defined by 
Duparc in his study of the Wadge hierarchy  \cite{dup}. 

\begin{Deff}[Duparc \cite{dup}]\label{til}
Let  $\Si$ be a finite alphabet  and $\tla  \notin \Si$, let $X=\Si\cup \{\tla\}$. 
Let $x$ be a finite or infinite word over the alphabet $X=\Si\cup \{\tla\}$. 
\nl Then  $x^\tla$ is inductively defined by:
\nl $\lambda^\tla =\lambda$,
\nl and for a finite word $u\in (\Si\cup \{\tla\})^\star$:  
\nl $(u.a)^\tla=u^\tla.a$, if $a\in \Si$,
\nl $(u.\tla)^\tla =u^\tla$  with its last letter removed if $|u^\tla|>0$,  
\nl i.e. $(u.\tla)^\tla =u^\tla(1).u^\tla(2)\ldots u^\tla(|u^\tla|-1)$  if $|u^\tla|>0$,
\nl $(u.\tla)^\tla=\lambda$  if $|u^\tla|=0$,
\nl and for $u$ infinite:
\nl $(u)^\tla = \lim_{n\in\om} (u[n])^\tla$, where, given $\beta_n$ and $v$ in   $\Si^\star$,
\nl $v\sqsubseteq \lim_{n\in\om} \beta_n \lra  \exists n \fa p\geq n\quad  \beta_p[|v|]=v$.
\nl(The finite {\it or} infinite word $\lim_{n\in\om} \beta_n$ is 
determined by the set of its (finite) prefixes).  
\end{Deff}

\begin{Rem}
For $x \in X^{\leq \om}$, $x^\tla$ denotes the string $x$, once every $\tla$ occuring in $x$
has been "evaluated" to the back space operation ( the one familiar to your computer!),
proceeding from left to right inside $x$. In other words $x^\tla = x$ from which every
 interval of the form $" a\tla "$ ($a\in \Si$) is removed.
\end{Rem}

\noi For example if $u=(a\tla)^n$, for $n$ an integer $\geq 1$, or 
$u=(a\tla)^\om$,  or $u=(a\tla\tla)^\om$,  then $(u)^\tla=\lambda$.
If $u=(ab\tla)^\om$ then $(u)^\tla=a^\om$  and 
 if $u=bb(\tla a)^\om$ then $(u)^\tla=b$. 

\hs Let us notice that in Definition \ref{til} the limit is not defined in the usual way: 
\nl for example if $u=bb(\tla a)^\om$ the finite word  $u[n]^\tla$ is alternatively 
equal to $b$ or to $ba$: more precisely $u[2n+1]^\tla=b$ and $u[2n+2]^\tla=ba$ for every 
integer $n\geq 1$ (it holds also that 
$u[1]^\tla=b$ and  $u[2]^\tla=bb$). Thus Definition \ref{til} implies that 
$\lim_{n\in\om} (u[n])^\tla = b$ so $u^\tla=b$. 

\hs  We can now define the operation $A \ra A^\sim$ of 
{\it exponentiation of conciliating sets}:

\begin{Deff}[Duparc \cite{dup}]
For $A\subseteq \Si^{\leq \om}$ and $\tla~  \notin \Si$, let $X=\Si\cup \{\tla\}$ and
\nl $A^\sim =_{df} \{x\in (\Si\cup \{\tla\})^{\leq \om} \mid  x^\tla\in A\}$.
\end{Deff}

\noi The operation $\sim$ is monotone with regard to the Wadge ordering and produce some sets
of higher complexity, as we shall see below. 
We shall need the notion of {\it cofinality 
of an ordinal} which may be found in \cite{sier,ck} and which we briefly recall now. 

\begin{Deff}
Let $\alpha$ be a limit ordinal, the cofinality of $\alpha$, denoted $cof(\alpha)$, 
is the least ordinal $\beta$ such that there exists a strictly increasing sequence of ordinals 
$(\alpha_i)_{i<\beta}$, of length $\beta$, such that 
for all $ i< \beta$,   ~ $\alpha_i < \alpha$,  ~ $\mbox{ and } $
$\sup_{i< \beta} \alpha_i = \alpha$. 
\noi This definition is usually extended to 0 and to the successor ordinals:
$cof(0)=0 \mbox{ and } cof(\alpha +1)=1 \mbox{ for every ordinal  } \alpha$. 
\end{Deff}

\noi The cofinality of a {\it limit ordinal} is always a limit ordinal satisfying:
$\om \leq cof(\alpha) \leq \alpha $. 
The ordinal  $cof(\alpha)$ is in fact a {\it cardinal} \cite{ck}.  
Then if the cofinality of a limit 
ordinal $\alpha$ is $\leq \om_1$, only the following cases may happen:
$cof(\alpha)=\om 
\mbox{     or      } cof(\alpha)=\om_1$.  
In this paper we shall not have to consider cofinalities which are larger than $\om_1$. 

 \hs  We can now state that the operation of exponentiation of 
conciliating sets is closely  related to ordinal exponentiation of base $\om_1$:

\begin{The} [Duparc \cite{dup}]\label{degexp}
Let $A\subseteq \Si^{\leq \om}$ be a conciliating set such that $A^d$ is a 
${\bf \Delta^0_\om }$-Borel set and 
$d_c(A)=d_w(A^d)=\alpha + n$ with $\alpha$ a limit ordinal and $n$ an integer $\geq 0$. Then 
$(A^\sim)^d$ is a ${\bf \Delta^0_\om }$-Borel set and there are three cases:
\begin{itemize}
\ite[a)]  If $\alpha = 0$, then $d_c( A^\sim )= (\om_1)^{d_c(A)-1}$
\ite[b)]   If $\alpha$ has cofinality $\om$, then $d_c( A^\sim )= (\om_1)^{d_c(A)+1}$
\ite[c)]   If $\alpha$ has cofinality $\om_1$, then $d_c( A^\sim )= (\om_1)^{d_c(A)}$
\end{itemize}
\end{The}

\noi Consider now this operation $\sim$ over infinitary context free  languages. 

\begin{The}[\cite{fi,fiwa}]\label{tildecf}
 Whenever $A\subseteq \Si^\om$ (respectively, $A\subseteq \Si^{\leq \om}$) is in $CFL_\om$,
(respectively, in $CFL_{\leq \om}$), then
 $A^\sim$ is in $CFL_\om$,
(respectively, in $CFL_{\leq \om}$). And $A, A^- \in CFL_{\leq \om}$ implies that 
$A^\sim, {(A^\sim)}^-={(A^-)}^\sim \in CFL_{\leq \om}$ 
\end{The}

\subsection{Operation of Iterated Exponentiation} 

\noi  In this section we are going to define a new operation $A \ra A^\bullet$ which 
can be called iterated exponentiation. It will involve an infinite number of erasers 
so each eraser will be coded  over a fixed finite alphabet and we shall see how a pushdown 
automaton will be able to guess, in a non deterministic way, 
that the erasing operations are correctly achieved in an input word.  

\hs  
One can already  iterate  the operation of exponentiation of sets. We shall use, in order 
to simplify our proofs, a variant $A^\approx$ of $A^\sim$ we already introduced 
in \cite{fi,f01}. $A^\approx$ is defined as $A^\sim$ with the only difference 
that in the definition \ref{til}, we write:  $(u.\tla)^\tla$ is undefined  if $|u^\tla|=0$, 
instead of $(u.\tla)^\tla=\lambda$  if $|u^\tla|=0$. Then one can show, as in \cite{fi},  
 that if $A\subseteq \Si^{\leq \om}$ and $d_c(A)\geq 2$ (hence $A^-\neq \emptyset$), 
then $A^\sim$ and $A^\approx$ are (conciliating) Wadge equivalent. 
 \nl  We define now, for a set $A\subseteq \Si^{\leq \om}$:
 $A^{\approx .0}=A$,  $A^{\approx .1}=A^\approx$  and  
$A^{\approx .(k+1)}=(A^{\approx .k})^\approx$, 
  where we apply $k+1$ times the operation 
$A\ra A^\approx$ 
with different new letters $\tla_1$, $\tla_2$, $\tla_3$, \ldots , $\tla_{k+1}$.  
  But this way, from a Borel conciliating set of 
finite rank,  we obtain only (conciliating)  Borel sets of finite ranks, of Wadge degree 
$< ^1\varepsilon_0$. A way to get sets of higher degrees, 
is  to use the {\it supremum of the sets} $A^{\approx .i}$. More generally we set the following 
definition.  

\begin{Deff}\label{defsup}
Let $\Si$ be a finite or countably infinite alphabet containing at least 
two letters $a$ and $b$ and let $(A_i)_{i\in \mathbb{N}}$ be a family of subsets of 
$\Si^{\leq \om}$. Then
$\sup_{i\in \mathbb{N}}A_i =_{df} \bigcup_{i\in \mathbb{N}} a^{i}.b.A_i$. 

\end{Deff}

\noi Let us recall now the following result. Notice that we give it in the general case 
of a countable (and possibly infinite) alphabet although we have only defined the conciliating 
Wadge hierarchy for conciliating sets defined over a {\it finite} alphabet 
(in order to simplify the presentation). 
\nl In fact the following proposition will only be used later in the case of a finite alphabet 
$\Si$, but we state it here (without details)   
in a more general case in order to give an indication of what we could obtain 
(all the $A^{\approx .i}$ are defined over the same {\it infinite} alphabet 
$\Si \cup \{\tla_1, \tla_2, \ldots \tla_k \ldots \}$). 

\begin{Pro}[\cite{dup}]\label{sup}
Let $\Si$ be a finite or countably infinite alphabet containing at least 
two letters $a$ and $b$ and let $(A_i)_{i\in \mathbb{N}}$ be a family of subsets of 
$\Si^{\leq \om}$ with $(A_i)^d$ Borel. Assume moreover  that  
$\fa i \in \mathbb{N}~~\exists  j_i \in \mathbb{N}~~ \mbox{ such that } d_c(A_i)<d_c(A_{j_i})$. 
 Then $(\sup_{i\in \mathbb{N}}A_i)^d$ is Borel and 
$d_c(\sup_{i\in \mathbb{N}}A_i) = \sup_{i\in \mathbb{N}} d_c(A_i)$. 

\end{Pro}

\noi Let us return now to the case of the supremum 
$\sup_{i\in \mathbb{N}}A^{\approx .i} = \bigcup_{i\in \mathbb{N}} a^{i}.b.A^{\approx .i}$ 
of the sets $A^{\approx .i}$.   It is defined over an infinite alphabet, and any  
infinitary context free language is defined over a finite alphabet. So we have  first to code 
this set over a finite alphabet. 
 The conciliating set $A^{\approx .n}$ is defined over the alphabet 
$\Si\cup\{\tla_1,\ldots , \tla_n\}$ hence we have to
 code every eraser $\tla_j$ by a finite word 
over a fixed finite alphabet. We shall code the eraser $\tla_j$ by the finite word 
$\alpha.B^j.C^j.D^j.E^j.\beta$ over the alphabet $\{\alpha, B, C, D, E, \beta\}$. 
We shall construct below a pushdown automaton accepting an infinitary language close 
to the coding of $\sup_{i\in \mathbb{N}}A^{\approx .i}$, for $A\in CFL_{\leq \om}$. 
It will need to  read four 
times the integer $j$ characterizing the eraser $\tla_j$ and this justifies our coding of 
the erasers. 
\nl  Remark first that the  morphism: 
\nl $F_n:~~  (\Si\cup \{\tla_1, \ldots , \tla_n\})^\star  \ra  
(\Si \cup \{\alpha, \beta, B, C, D, E\})^\star$
\nl defined  by  $F_n(c)=c$ for each $c\in \Si$ and $F_n(\tla_j)=\alpha.B^j.C^j.D^j.E^j.\beta$ 
for each integer 
$j \in [1, n]$, where $B, C, D, E, \alpha, \beta$ are new letters not in $\Si$, can be   
 naturally extended  to a function: 
\nl $\bar{F_n}:~~  (\Si\cup \{\tla_1, \ldots , \tla_n\})^{\leq \om}  \ra 
 (\Si \cup \{\alpha, \beta, B, C, D, E\})^{\leq \om}$. 
\nl Using Wadge games, we can now state the following lemma. 

\begin{Lem}\label{4-22}
Let $A\subseteq \Si^{\leq \om}$ be such that $A^d$ is a ${\bf \Delta^0_\om }$-Borel set 
and $d_c(A)\geq 2$.  Then 
 $d_c(\bar{F_n}(A^{\approx .n}))=d_c(A^{\approx .n})$  holds for every  
integer $n\geq 2$. If moreover  
$\fa n \geq 1~~  d_c(A^{\approx .n}) < d_c(A^{\approx .(n+1)})$ then 
$d_c(\sup_{i\geq 1}\bar{F_i}(A^{\approx .i}))= 
\sup_{i\geq 1}d_c(A^{\approx .i})$. 
\end{Lem}

\noi We would like now to apply the above lemma to construct, from an infinitary 
context free language $A$ such that  $A^d$ is a ${\bf \Delta^0_\om }$-Borel set 
and $d_c(A)\geq 2$, another   infinitary 
context free language of Wadge degree   $\sup_{i\geq 1}d_c(A^{\approx .i})$. 
\nl But we can not show that, whenever $A\in CFL_{\leq \om}$, then 
 $\sup_{n\geq 1}\bar{F_n}(A^{\approx .n})$
is in $CFL_{\leq \om}$. This is connected to 
the fact that the finitary language 
$\{B^jC^jD^jE^j~\mid  j\geq 1\}$
is not a context free language. But its complement is easily seen to be context free. 
Then we shall sligthly modify the set  $\sup_{n\geq 1}\bar{F_n}(A^{\approx .n})$, 
in the following way. 
We can add to this language  all $(\leq \om)$-words in the form $a^n.b.u$ 
where there is in $u$ 
a segment $\alpha.B^j.C^k.D^l.E^m.\beta$, with $j, k, l, m$ integers $\geq 1$, 
which does not code any eraser, or codes an eraser $\tla_j$ for $j>n$. 

\hs  Define first the following context free finitary languages over the alphabet 
\nl $X^\square=(\Si \cup \{\alpha, \beta, B, C, D, E\}):$
\nl $L^B = \{ a^n.b.u.B^j~ \mid ~ 
n\geq 1 \mbox{ and } j>n \mbox{ and } u\in (X^\square)^\star\}$
\nl $L^C = \{ a^n.b.u.C^j~ \mid ~ 
n\geq 1 \mbox{ and } j>n \mbox{ and } u\in (X^\square)^\star\}$
\nl $L^D = \{ a^n.b.u.D^j~ \mid ~
 n\geq 1 \mbox{ and } j>n \mbox{ and } u\in (X^\square)^\star\}$
\nl $L^E = \{ a^n.b.u.E^j~ \mid ~ 
n\geq 1 \mbox{ and } j>n \mbox{ and } u\in (X^\square)^\star\}$
\nl $L^{(B,C)} = \{ u.\alpha.B^j.C^k.D^l.E^m.\beta ~ \mid ~ j, k, l, m\geq 1 
\mbox{ and } j\neq k \mbox{ and } u\in a^+.b.(X^\square)^\star\}$
\nl $L^{(C,D)} = \{ u.\alpha.B^j.C^k.D^l.E^m.\beta ~ \mid ~ j, k, l, m\geq 1 
\mbox{ and } k\neq l \mbox{ and } u\in a^+.b.(X^\square)^\star\}$
\nl $L^{(D,E)} = \{ u.\alpha.B^j.C^k.D^l.E^m.\beta ~ \mid ~ j, k, l, m\geq 1 
\mbox{ and } l\neq m \mbox{ and } u\in a^+.b.(X^\square)^\star\}$

\hs  It is easy to show that each of these languages is a context 
free finitary language thus 
$L = L^B \cup L^C  \cup L^D \cup L^E \cup L^{(B,C)} 
\cup L^{(C,D)} \cup L^{(D,E)}$ is also   context free because the class 
CFL is closed under finite union. Then $L.(X^\square)^{\leq \om}$ is an 
infinitary CFL.
 Remark  that all words in $\sup_{n\geq 1}\bar{F_n}(A^{\approx .n})$ belong to 
the infinitary regular language 
$R = a^+.b.(\Si \cup (\alpha.B^+.C^+.D^+.E^+.\beta))^{\leq \om}$.
 Consider now the language
$L.(X^\square)^{\leq \om}  \cap R$. 
 A word $\sigma$ in this language is a word in $R$,  
having an initial segment in the form $a^n.b$, with $n\geq 1$, and  
 containing a segment $\alpha.B^j.C^k.d^l.E^m.\beta$ with $j, k, l, m \geq 1$ 
which does not code any eraser $\tla_i$ or codes such an eraser but with $i>n$. 
Define now 
$$A^\bullet = \sup_{n\geq 1}\bar{F_n}(A^{\approx .n}) \cup 
[ L.(X^\square)^{\leq \om}  \cap R ]$$

\noi We shall show that the operation $A \ra A^\bullet$ conserves the context freeness of 
infinitary languages and that if $A$ is a ${\bf \Delta^0_\om }$-set of Wadge degree $\geq 2$ 
then $A^\bullet$ and $\sup_{n\geq 1}\bar{F_n}(A^{\approx .n})$ are Wadge equivalent. 
So we shall be able to construct, 
from such an infinitary context free language $A$, 
another infinitary context free language 
of Wadge degree $\sup_{i\geq 1}d_c(A^{\approx .i})$.

\begin{The}\label{the}
 If   $A\subseteq \Si^{\leq \om}$ is  an infinitary context free language 
then $A^\bullet$ is an infinitary context free language over the alphabet 
$X^\square=(\Si \cup \{\alpha, \beta, B, C, D, E\})$.   
\end{The}

\proo It relies on 
a technical construction of a pushdown automaton accepting $A^\bullet$ from   
a pushdown automaton accepting $A$. The idea of the construction is already in 
\cite{f01}, where $A$ was assumed to be an \orl~ and where we  proved only 
the existence of some $\om$-context free languages which are Borel sets of infinite rank. 
We shall give here a similar 
 construction in the more general case of an infinitary context free 
language $A$. 

\hs Let then $A \subseteq \Si^{\leq \om}$ be an infinitary context free language 
($A$ may contain {\it finite and infinite words}).  
We can write $A = A_1 \cup A_2$ where  
$A_1 \subseteq \Si^{\om}$ is an $\om$-CFL and $A_2 \subseteq \Si^\star$ is a 
finitary context free language.  Then $A^\bullet = A_1^\bullet \cup A_2^\bullet$ 
holds by definition of  $A^\bullet$. 
\nl The $\om$-language  $A_1\subseteq \Si^{\om}$ is accepted by 
a Muller pushdown automaton $\mathcal{A}_1$.
Remark that in that case all  sets  $A_1^{\approx .n}$ as well as 
$\sup_{n\geq 1}\bar{F_n}(A_1^{\approx .n})$ contain only infinite words.  
       
\hs We shall  find a MPDA $\mathcal{B}$ accepting an $\om$-CFL $L(\mathcal{B})$ 
such that 
$$\sup_{n\geq 1}\bar{F_n}(A_1^{\approx .n}) \subseteq L(\mathcal{B}) \subseteq 
A_1^\bullet = \sup_{n\geq 1}\bar{F_n}(A_1^{\approx .n}) 
\cup [ L.(X^\square)^{\leq \om}  \cap R ]$$ 

\noi Thus we shall have 
$A_1^\bullet = L(\mathcal{B}) \cup [ L.(X^\square)^{\leq \om}  \cap R ]$ 
and this will imply that $A_1^\bullet$ is in $CFL_{\leq \om}$  because the class 
$CFL_{\leq \om}$   is closed under finite union. 
\nl   It is easy to have  $L(\mathcal{B}) \subseteq R$ because if $L(\mathcal{B'})$ is 
an $\om$-CFL which is not included into $R$ one can  replace it by 
 $L(\mathcal{B})=L(\mathcal{B'}) \cap R $ which is then an $\om$-CFL verifying 
$L(\mathcal{B})=L(\mathcal{B'}) \cap R \subseteq R$.
\nl  Recall  now that 
$L.(X^\square)^{\leq \om}  \cap R$
\noi is the set of {\bf {\it all} } (finite or infinite) words in $R$ but not in 
$\cup_{n\geq 1}a^n.b.\bar{F_n}(\Si\cup \{\tla_1, \ldots , \tla_n\})^{\leq \om}$. 
\nl Thus,  in order to define the MPDA $\mathcal{B}$, we have  only to 
 consider   the behaviour of $\mathcal{B}$ when reading $\om$-words in 
$\cup_{n\geq 1}a^n.b.\bar{F_n}(\Si\cup \{\tla_1, \ldots , \tla_n\})^{\om}$ 
and  we have to find  a MPDA $\mathcal{B}$ 
such that $L(\mathcal{B})$ contains  such a word $a^n.b.u$ if and only if  
$u\in \bar{F_n}(A_1^{\approx .n})$. 

\hs  So we have to look first at $\om$-words in $\bar{F_n}(A_1^{\approx .n})$.  In such a 
word $u \in \bar{F_n}(A_1^{\approx .n})$, there are (codes of) erasers $\tla_1, \ldots, 
\tla_n$. In order to simplify our notations, we shall sometimes write in the sequel 
$\tla_j=\alpha.B^j.C^j.D^j.E^j.\beta$ and call eraser either $\tla_j$ or its code 
$\alpha.B^j.C^j.D^j.E^j.\beta$. 
The $\om$-word $u$ is in $\bar{F_n}(A_1^{\approx .n})$ if and only if 
after  the operations of erasing ( with the erasers 
$\tla_1,..., \tla_n$ ) have been achieved in $u$, then the resulting word 
is in $A_1$.
\nl  Because of the inductive definition of the sets $A_1^{\approx .n}$, the operations 
of erasing have to be  done in a good order:  in an $\om$-word 
which contains only the erasers $\tla_1,..., \tla_n$, the first operation of erasing uses the 
last eraser $\tla_n$, then the second one uses the eraser $\tla_{n-1}$, and so on \ldots 
\nl  We  now informally describe the behaviour of the MPDA  $\mathcal{B}$ when reading 
an $\om$-word $a^n.b.u$ 
with $u \in  \bar{F_n}(\Si\cup \{\tla_1, \ldots , \tla_n\})^{\om}$. The MPDA $\mathcal{B}$ 
will generalize the MPDA accepting $A_1^\approx$ constructed in \cite{fi}. 
\nl  After the reading of the  initial segment in the form $a^n.b$, the MPDA  $\mathcal{B}$ 
simulates the MPDA  $\mathcal{A}_1$ until it guesses, using the non 
determinism, that it begins to read a segment 
$w$ which contains erasers which really erase and some letters of $\Si$ or some other 
erasers which are erased when the operations 
of erasing are achieved in $u$. 
\nl Then,  using the non determinism, when   $\mathcal{B}$ reads 
a letter $c\in \Si$
and  guesses that this letter will be erased it pushes it in the pushdown store, 
keeping in memory the current global state (consisting in the stack content 
and the current state of the finite control) of the MPDA $\mathcal{A}_1$. 
\nl  In a similar manner, when   $\mathcal{B}$ reads 
the  code $\tla_j=\alpha.B^j.C^j.D^j.E^j.\beta$ of an eraser and  guesses that 
this eraser will be erased (by another eraser $\tla_k$ with $k>j$),  it pushes 
in the store the finite word $\gamma.E^j.\varepsilon$ (coding the eraser $\tla_j$ 
in the stack), where $\gamma, \varepsilon$ are 
in the stack alphabet of $\mathcal{B}$. 
\nl  But  $\mathcal{B}$ 
may also guess that the eraser $\tla_j=\alpha.B^j.C^j.D^j.E^j.\beta$ will really be used
as an eraser.  In that case
$\mathcal{B}$ has to pop from the top of the pushdown store either a letter  $c\in \Si$ or 
the code $\gamma.E^i.\varepsilon$ of another eraser $\tla_i$, with $i<j$, which 
is erased by $\tla_j$. 
\nl It would be  easy for $\mathcal{B}$ to check whether $i<j$ when reading the initial segment 
$\alpha.B^j$ of $\tla_j$. 
\nl  The pushdown $\mathcal{B}$ has also to ensure that 
the operations of erasing are achieved in a good order. This can be done, using our 
coding of erasers containing four times the integer $j$ characterizing $\tla_j$.      
We refer to \cite{f01} where this behaviour of $\mathcal{B}$ is described. 

\hs Consider now $A_2\subseteq \Si^{\star}$. 
Then $\sup_{n\geq 1}\bar{F_n}(A_2^{\approx .n})$  may contain 
 {\it finite and infinite} 
words.  Then $\sup_{n\geq 1}\bar{F_n}(A_2^{\approx .n})=L_1 \cup L_2$ where 
$L_1$ is a finitary language and $L_2$ is an \ol~ over the alphabet 
$X^\square$. 
 Following the same ideas as in the preceding case (where $A_1\subseteq \Si^{\om}$) we can 
construct a   pushdown automaton  
$\mathcal{B}'$ accepting a finitary context free language $L(\mathcal{B}')$
and a Muller pushdown automaton   $\mathcal{B}''$ accepting an $\om$-CFL 
$L(\mathcal{B}'')$   such that:
$$L_1  \subseteq  L(\mathcal{B}') \subseteq L_1 \cup [ L.(X^\square)^{\leq \om}  \cap R ]$$ 
$$L_2  \subseteq  L(\mathcal{B}'') \subseteq L_2 \cup [ L.(X^\square)^{\leq \om}  \cap R ]$$ 
\noi Then it turns out that                    
 $A_2^\bullet = L(\mathcal{B}') \cup 
L(\mathcal{B}'') \cup [ L.(X^\square)^{\leq \om}  \cap R ]$ is  in $CFL_{\leq \om}$. 

 \hs Consider now again 
 the infinitary context free language  $A \subseteq \Si^{\leq \om}$.   
It turns out that  $A^\bullet = A_1^\bullet \cup A_2^\bullet$ 
 is an infinitary context free language because the class $CFL_{\leq \om}$  
 is closed under finite union. 
\ep 
  
\hs  The operation $A \ra A^\bullet$ will provide a kind of infinite iteration of the operation 
$A \ra A^\sim$. Thus the above theorem will enable us 
to  get some infinitary 
context free languages of larger Wadge degrees than those we could previously obtain. 

\hs 
In order to give precisely the Wadge degree of $A^\bullet$ from the Wadge degree of $A$ we 
introduce now some notations for ordinals.  For an ordinal $\alpha$ we  define 
$\om_1(1, \alpha)=\om_1^\alpha$ and  for an integer $n\geq 1$, 
 $\om_1(n+1, \alpha)=\om_1^{\om_1(n, \alpha)}$. 
 If $\alpha \leq$ $^1\varepsilon_0$  the limit of the sequence of 
ordinals $\om_1(n, \alpha)$ 
is  the  ordinal $^1\varepsilon_0$. 
And if $\alpha >$ $^1\varepsilon_0$  the limit of the sequence of  ordinals $\om_1(n, \alpha)$ 
is the  first fixed point 
of the operation of ordinal 
exponentiation of base $\om_1$ which is greater than (or equal to) $\alpha$. 
We shall denote it  $^1\varepsilon_0(\alpha)$. 
 Then one can enumerate the sequence of the $\om$ first fixed points 
of the operation $\alpha \ra \om_1^\alpha$, which are:
 $^1\varepsilon_0$,  
 $^1\varepsilon_1 = \,^1\varepsilon_0 (^1\varepsilon_0 + 1)$,  
 $^1\varepsilon_2 = \, ^1\varepsilon_0 (^1\varepsilon_1 + 1)$, 
 and for each integer $n\geq 0$:
$^1\varepsilon_{n+1} = \, ^1\varepsilon_0 (^1\varepsilon_n + 1)$. 
 The next fixed point is the $\om^{th}$ fixed point, denoted $^1\varepsilon_\om$, 
and it is also the limit of the sequence of fixed points $^1\varepsilon_n$, for $n\geq 0$:
 $^1\varepsilon_\om = \sup_{n\in \om} (^1\varepsilon_n)$.
 The sequence of fixed points of the operation of exponentiation of base $\om_1$ continues 
beyond this ordinal because, for each ordinal $\alpha$, 
 there exists such a fixed point which is greater than $\alpha$. 
These  fixed points are indexed by ordinals and they are defined by induction on the ordinals. 
For every  successor ordinal $\beta + 1$, the ordinal $^1\varepsilon_{\beta +1}$ is 
defined as above by: 
$^1\varepsilon_{\beta +1} = \, ^1\varepsilon_0 (^1\varepsilon_\beta + 1)$. 
And for a limit ordinal $\delta$ the ordinal  $^1\varepsilon_{\delta}$ is defined by 
$^1\varepsilon_{\delta} = \sup_{\beta < \delta} (^1\varepsilon_\beta)$.
 
\hs
If $A$ is a ${\bf \Delta^0_\om }$-Borel set then its  Wadge degree is smaller than the ordinal 
$\,^1\varepsilon_{\om_1}$ which is  
the $(\om_1)^{th}$ fixed point of the operation of ordinal exponentiation 
of base $\om_1$. 
But for all ordinals $\delta < \om_1$ the ordinal $^1\varepsilon_{\delta}$ has cofinality 
smaller than $\om_1$.  
Thus   $d_w(A)$ cannot be a fixed point of cofinality $\om_1$ of the 
operation of ordinal exponentiation of base $\om_1$. The following Proposition easily 
follows from this fact and from Theorem \ref{degexp}.

\begin{Pro}
Let $A\subseteq \Si^{\leq \om}$ be a conciliating set such that $A^d$ is a 
${\bf \Delta^0_\om }$-Borel set and $d_c(A)\geq 2$. Then $d_c( A^\sim ) > d_c(A)$. 
\end{Pro}

\begin{Rem}  
Let $A\subseteq \Si^{\leq \om}$ be such that $A^d$ is a ${\bf \Delta^0_\om }$-Borel set 
and $d_c(A)\geq 2$.  Then one can easily show by induction  that 
$\fa n \geq 1~~  d_c(A^{\approx .n}) < d_c(A^{\approx .(n+1)})$.  
So this  additional hypothesis we have made in Lemma \ref{4-22}
is in fact always verified.    
\end{Rem}

\noi The counterpart of  the operation $A \ra A^\bullet$ with regard to 
 Wadge degrees is given precisely 
by the following theorem.

\begin{The}\label{A-bul-inf}
\footnote{In the short version of this paper which appeared in the 
proceedings of LCCS 01 we omitted the distinction between 
items (1) and (2) of this theorem.}
Let  $A\subseteq \Si^{\leq \om}$ be  such that $A^d$ is a ${\bf \Delta^0_\om }$-Borel set 
and $d_c(A)\geq 2$. 
\begin{enumerate}
\ite[(1)] 
If $d_c(A)$ is  a fixed point (of cofinality $\om$) of the operation 
of exponentiation of base $\om_1$: $\alpha \ra \om_1^\alpha$,  then 
$d_c(A^\bullet)=\,^1\varepsilon_0(d_c(A)+1)$.
\ite[(2)] If  
$d_c(A)$ is not a fixed point  of the operation $\alpha \ra \om_1^\alpha$, 
then $d_c(A^\bullet)=\,^1\varepsilon_0(d_c(A))$.
\ite[(3)] 
 $d_c(A^\bullet)$ is the first fixed point of this operation 
which is strictly larger than $d_c(A)$ 
and $(A^\bullet)^- \equiv_c A^\bullet \cup a^{\leq \om}$ holds. 

\end{enumerate}
\end{The}

\proo Let $A\subseteq \Si^{\leq \om}$ be such that $A^d$ is a 
${\bf \Delta^0_\om }$-Borel set and $d_c(A)\geq 2$.
We can prove that $A^\bullet$ and $\sup_{n\geq 1}\bar{F_n}(A^{\approx .n})$ are 
conciliating Wadge equivalent, using the conciliating Wadge game, and examining in detail 
several cases which can happen.  
\nl Then items (1) and (2) of 
Theorem \ref{A-bul-inf} can be easily derived  from Theorem \ref{degexp} and Proposition 
\ref{sup}.
It follows from items (1) and (2) that  $d_c(A^\bullet)$ is the first fixed point 
of the operation of ordinal 
exponentiation of base $\om_1$ 
which is strictly larger than $d_c(A)$.  
\nl We can now prove that player 1 has a 
winning strategy in the conciliating Wadge game $C(A^\bullet, A^\bullet \cup a^{\leq \om})$ 
and also in the conciliating Wadge game $C(A^\bullet \cup a^{\leq \om}, A^\bullet)$. 
This implies that neither $A^\bullet \cup a^{\leq \om} \leq_c  A^\bullet$ nor 
$A^\bullet \leq_c  A^\bullet \cup a^{\leq \om}$ hold.  Then it follows from the properties 
of the conciliating hierarchy that
$(A^\bullet)^- \equiv_c A^\bullet \cup a^{\leq \om}$.  
\ep 

\hs
If $A$ is an infinitary context free language then 
$A^\bullet$ and $A^\bullet \cup a^{\leq \om}$ are infinitary context free languages. 
So if moreover  $A^d$ is a ${\bf \Delta^0_\om }$-Borel set 
and $d_c(A)\geq 2$ then 
there  exists an infinitary context free language which is  conciliating Wadge equivalent 
to $(A^\bullet)^-$. This fact will be useful in next   section. 
\nl 
Remark also that in particular if $2 \leq d_c(A) <^1\varepsilon_0$, i.e. if $A^d$ is  Borel 
of finite rank and of Wadge degree $\geq 2$, then  $d_c(A^\bullet)=^1\varepsilon_0$, and  
$A^\bullet$ is a Borel set of rank $\om$. 

\section{Wadge Hierarchy of Infinitary Context Free Languages}

\noi 
If we consider the operation of ordinal exponentiation of base $\om$: $\alpha \ra \om^\alpha$, 
one can define in a similar way as above the successive fixed points of this 
operation. These ordinals are the well known Cantor ordinals $\varepsilon_0$, 
$\varepsilon_1$, \ldots and $\varepsilon_\om$ is the $\om^{th}$ such fixed point, \cite{sier}.  
\nl From the above closure properties of the class $CFL_{\leq \om}$ 
under the operations of sum, of exponentiation and of iterated exponentiation, and using 
the correspondence  between these operations and the arithmetical operations 
over ordinals, one can show the following:

\begin{The}\label{thewa}
The length of the conciliating hierarchy of infinitary languages in 
$CFL_{\leq \om}$ is  greater  than or equal to $\varepsilon_\om$. 
 The length of the Wadge hierarchy of $\om$-context free languages  in 
$CFL_{\om} \cap {\bf \Delta^0_\om }$ 
 is  greater  than or equal to $\varepsilon_\om $. 
\end{The}

 \proo We firstly define  a strictly increasing function $H$ from $\varepsilon_\om$ into 
$^1\varepsilon_\om$.  This function is defined as follows: first 
$H(n)=n$ ~~ for each integer  $n$
and $H(\varepsilon_i) =\,^1\varepsilon_i$~~ for each integer $i\geq 0$. 
 Next if $\alpha$ is a non null ordinal $< \varepsilon_\om$, it has an 
iterated Cantor normal form of base $\om$ \cite{sier}: 
$$\alpha = \om^{\alpha_j}.m_j + \om^{\alpha_{j-1}}.m_{j-1} + \ldots  +  \om^{\alpha_1}.m_1$$

\hs  where  $j>0$ is an integer,
$\varepsilon_\om > \alpha \geq \alpha_j > \alpha_{j-1} > \ldots > \alpha_1 $  are ordinals and 
$m_j, m_{j-1}, \ldots , m_1$ are integers  $ > 0$. 
And where each $\alpha_i$ itself is written  in Cantor normal form of base $\om$, 
and so on. Then one can inductively define $H'(\alpha)$ and $H(\alpha)$ in the following way. 

\hs We first set 
$$H'(\alpha) = \om_1^{H(\alpha_j)}.m_j + \om_1^{H(\alpha_{j-1})}.m_{j-1} + \ldots  
+  \om_1^{H(\alpha_1)}.m_1$$

\noi and we distinguish now two cases: 

\hs {\bf First case.}  $H'(\alpha)=\beta + n$ with $\beta$ a limit ordinal of 
cofinality $\om$, $\beta \neq \,^1\varepsilon_i$ for all integers $i\geq 0$, and 
$n$ an integer $\geq 0$. 
\nl In that case we set $H(\alpha)=H'(\alpha) + 1$.

\hs {\bf Second case.} $H'(\alpha)=\beta + n$  with $\beta$ a limit ordinal of 
cofinality $\om_1$ or $\beta = \,^1\varepsilon_i$ for some integer $i \geq 0$, and 
$n$ an integer $\geq 0$.  
\nl In that case we set $H(\alpha)=H'(\alpha)$.

\hs So the shift we introduce in the first case is used to avoid the ordinal $H(\alpha)$
to be a limit ordinal of 
cofinality $\om$, different from  $\,^1\varepsilon_i$ for all integers $i\geq 0$,  
while  the function $H$ remains  strictly increasing. 

\hs Let us give now some examples. For ~~
$\alpha = \varepsilon_2  + 4$, ~~
 the above definition leads to 
\nl ~~$H(\alpha) = \,^1\varepsilon_2 + 4$,  
while for  
~~$\alpha = \varepsilon_2  + \varepsilon_1 + 4$~~
it holds that ~~
$H(\alpha) = \,^1\varepsilon_2 + \,^1\varepsilon_1  + 5$.  
\nl  For ~~
$\alpha = \varepsilon_2.3 + \om^{(\varepsilon_1 + \om^\om)} + \om^{(\om^\om +2)} 
$,~~
\noi the above definition leads to 
\nl $H(\alpha) = \,^1\varepsilon_2.3 + \om_1^{(^1\varepsilon_1 + \om_1^{\om_1})} + 
\om_1^{(\om_1^{\om_1} +2)}$. 

\noi  and for 
$$\alpha = \varepsilon_4.3 + \om^{(\varepsilon_3 + \varepsilon_1 )} 
+ \om^{(\varepsilon_2 + \varepsilon_1 + 5)} + \varepsilon_2 + 3$$ 
\noi it holds that 
$$H(\alpha) = \,^1\varepsilon_4.3  + \om_1^{(\,^1\varepsilon_3 + \,^1\varepsilon_1 +1)} 
 + \om_1^{(\,^1\varepsilon_2 + \,^1\varepsilon_1 + 6)} + \,^1\varepsilon_2 + 4$$

\noi It is easy to show that the function $H$ is stricly increasing, thus 
the image $H(\varepsilon_\om )$ is of order type $\varepsilon_\om$, and so is 
$H(\varepsilon_\om )-\{0\}$. 

\hs We can now follow Definition 32 of \cite{dup} and define a conciliating context free language 
$\Omega(H(\alpha))$  of degree $H(\alpha)$, for each non null ordinal 
$\alpha <  \varepsilon_\om $. 
\nl We shall need also to define a conciliating context free language 
$\Omega(H'(\alpha))$  of degree $H'(\alpha)$ in the case $H'(\alpha)$ 
is an ordinal of cofinality $\om$ different from $ ^1\varepsilon_i$ for all 
integers $i\geq 0$. 
\nl  Let $\delta$ be a non null  ordinal in $H(\varepsilon_\om )$. Then  
$\delta < \,^1\varepsilon_\om $ hence  $\delta$ admits an iterated Cantor normal form of base 
$\om_1$, \cite{sier}:

$$\delta  = \om_1^{\delta _j}.\nu_j + \om_1^{\delta _{j-1}}.\nu_{j-1} + \ldots  +  
\om_1^{\delta _1}.\nu_1$$

\noi where  $j>0$ is an integer,
$^1\varepsilon_\om > \delta \geq \delta _j > \delta _{j-1} > \ldots > \delta _1 $  
are ordinals and 
$\nu_j, \nu_{j-1}, \ldots , \nu_1$ are non null ordinals $< \om_1$, 
and  each $\delta_i$ itself is written  in Cantor normal form of base $\om_1$, 
and so on \ldots 
\nl But here {\bf each ordinal $\nu_i$ is an integer} because 
$\delta \in H(\varepsilon_\om )$ and for each $i$, $\delta _i\in H(\varepsilon_\om )$ also 
holds. Then one can inductively define the set  
$$\Omega(\delta) = \Omega(\om_1^{\delta _j}).\nu_j + 
\Omega(\om_1^{\delta _{j-1}}).\nu_{j-1} + \ldots  + \Omega(\om_1^{\delta _1}).\nu_1$$

\noi where $\Omega(\om_1^{\beta})$ with $\beta <\, ^1\varepsilon_\om$ and 
$\beta \in H(\varepsilon_\om)$ 
is defined by:

\begin{enumerate}
\ite[a)]  If $\beta =0$, then $\Omega(\om_1^{\beta})=\Omega(1)=\emptyset$. 
\ite[b)]   If $\beta =n > 0 $ is an integer, 
then $\Omega(\om_1^{\beta})=\Omega(\beta +1)^\sim$.
\ite[c)]  If $\beta =\gamma + n$ where $\gamma$ is an ordinal of cofinality 
$\om_1$ and $n$ is an integer $\geq 0$, then $\Omega(\om_1^{\beta})=\Omega(\beta)^\sim$.
\ite[d)]  If $\beta =\om_1^\beta=\,^1\varepsilon_i$, for some integer $i$, $0\leq i < \om$. 
\nl   We shall  construct some infinitary context free languages of (conciliating) 
Wadge degrees $^1\varepsilon_i$, for $0\leq i < \om$.  
Let  $A\in CFL_{\leq \om}$  such that  $ d_c( A)=2$ 
(for example $A=\emptyset + \emptyset$ where $\emptyset$ is the empty 
conciliating set,which is of Wadge degree $1$) 
then  $d_c(A^{\bullet}) = \,^1\varepsilon_0( d_c(A))=\,^1\varepsilon_0$. Denote  
$\Omega(^1\varepsilon_0) = A^{\bullet}$. The ordinal  $d_c(A^{\bullet}) = \,^1\varepsilon_0$ 
is a fixed point of the operation of exponentiation of base $\om_1$ which  has cofinality $\om$.  
Thus if $A^{\bullet.2}= (A^{\bullet})^{\bullet}$ then 
$$d_c(A^{\bullet.2})= \,^1\varepsilon_0( d_c(A^{\bullet}) +1) )= 
\,^1\varepsilon_0( ^1\varepsilon_0 + 1) = \,^1\varepsilon_1$$ 

\noi by Theorem \ref{A-bul-inf} (1). Next we can iterate this construction, 
defining inductively for integers $j\geq 2$ 
the sets $A^{\bullet.j}= (A^{\bullet.(j-1)})^{\bullet}$. Then one can prove by 
induction that for each integer $j\geq 2$  

$$d_c(A^{\bullet.j})= \,^1\varepsilon_0( d_c(A^{\bullet.(j-1)}) +1) )= 
\,^1\varepsilon_0( ^1\varepsilon_{j-2} + 1) = \,^1\varepsilon_{j-1}$$ 

\noi  Then we denote $\Omega(^1\varepsilon_i) = A^{\bullet.(i+1)}$ and 
$d_c(\Omega(^1\varepsilon_{i}))=\,^1\varepsilon_{i}$ holds for every integer $i\geq 0$.  
 \nl Remark that the above   construction is a particular case of Duparc's construction of a conciliating 
set $\Omega(\om_1^\beta)$ of degree $\om_1^\beta$ when $\beta$ is an ordinal of cofinality 
$\om$. Indeed the ordinals  $^1\varepsilon_{j}$, for  $0\leq j < \om$, have cofinality 
$\om$ because $^1\varepsilon_{0}=\sup_{n\geq 1}\om_1(n, 2)$ and  
$^1\varepsilon_{j+1}=\sup_{n\geq 1}\om_1(n, \,^1\varepsilon_j +1)$ holds for every integer 
$j\geq 0$. 

\ite[e)] If $\beta =\,^1\varepsilon_i + n$ where $i$ is an integer $\geq 0$ and 
$n$ is an integer $> 0$,  
then $\Omega(\om_1^{\beta})=\Omega(\beta - 1)^\sim$.

\ite[f)]  If $\beta = \gamma + n$ where $\gamma$ is an ordinal of cofinality 
$\om$, $n$ is an integer $> 1$, and $\gamma \neq ^1\varepsilon_i$ for all 
integers $i\geq 0$,  then $\Omega(\om_1^{\beta})=\Omega(\beta - 1)^\sim$.

\ite[g)]  If $\beta = \gamma + 1$ where $\gamma$ is an ordinal of cofinality 
$\om$, and $\gamma \neq ^1\varepsilon_i$ for all 
integers $i\geq 0$. 
\nl In that case there exists an ordinal $\alpha < \varepsilon_\omega$ such that 
$\gamma = H'(\alpha)$ and $\beta = H(\alpha) = H'(\alpha) +1$. 
\nl The Cantor normal form of base $\om$ of the ordinal $\alpha$ is:
$$\alpha = \om^{\alpha_j}.m_j + \om^{\alpha_{j-1}}.m_{j-1} + \ldots  +  \om^{\alpha_1}.m_1$$
where  $j>0$ is an integer,
$\varepsilon_\om > \alpha \geq \alpha_j > \alpha_{j-1} > \ldots > \alpha_1 $  are ordinals and 
$m_j, m_{j-1}, \ldots , m_1$ are integers  $ > 0$. 
\nl And  we have defined $H'(\alpha)$ by 
$$H'(\alpha) = \om_1^{H(\alpha_j)}.m_j + \om_1^{H(\alpha_{j-1})}.m_{j-1} + \ldots  
+  \om_1^{H(\alpha_1)}.m_1$$
\noi so we can inductively define 
$$\Omega(H'(\alpha)) = \Omega(\om_1^{H(\alpha_j)}).m_j + 
\Omega(\om_1^{H(\alpha_{j-1})}).m_{j-1} + \ldots  + \Omega(\om_1^{H(\alpha_1)}).m_1$$
\noi and we set 
$$\Omega(\om_1^{\beta})=\Omega(\beta - 1)^\sim =
 \Omega(\gamma)^\sim  = \Omega(H'(\alpha))^\sim $$

\ite[h)]  Notice that it is not necessary to define 
$\Omega(\om_1^{\beta})$ in the case of an ordinal  
$\beta$  of cofinality 
$\om$ and different from $ ^1\varepsilon_i$ for all 
integers $i\geq 0$. 
This case cannot happen here because of the shift we introduced in the first case of the 
definition of the ordinal $H(\alpha)$.  

\end{enumerate}

\noi The closure properties of the class $CFL_{\leq \om}$ under 
the operations of sum, of exponentiation and of iterated exponentiation imply that,  
for every ordinal $\delta \in  H(\varepsilon_\om )$, 
$\Omega(\delta) \in CFL_{\leq \om}$ holds 
and the complement of $\Omega(\delta)$ is (conciliating) Wadge equivalent 
to some conciliating set in $CFL_{\leq \om}$.

\hs By our construction  and by Theorems \ref{thesum}, 
\ref{degexp} and \ref{A-bul-inf} (or using Theorem  33 of \cite{dup}) 
$d_c(\Omega(\delta))=\delta < \,^1\varepsilon_\om$ holds 
for every  $\delta \in  H(\varepsilon_\om )$, hence  the length of the conciliating hierarchy 
of infinitary context free languages is greater than or equal to $\varepsilon_\om$. 

\hs We consider now the Wadge hierarchy of $\om$-context free languages. 
For each non null ordinal  $\alpha < \varepsilon_\om$, the \ol~ 
$\Omega(H(\alpha ))^d$ is a Borel set in the class 
$CFL_{\om}\cap {\bf \Delta^0_\om }$ and 
$d_w( \Omega(H(\alpha ))^d ) = d_c(\Omega(H(\alpha )))= H(\alpha)$. 

\hs This  proves that the length 
of the Wadge hierarchy of $\om$-context free languages in 
$CFL_{\om}\cap {\bf \Delta^0_\om }$ 
 is greater than or equal to the 
ordinal $\varepsilon_\om$. 
\ep

\section{${\bf \Si^0_\om }$-Complete $\om$-Context Free Language}

\noi 
With the operations we have studied above, one cannot reach, from a (conciliating) Borel set 
of finite rank,  a Borel set of  
Wadge degree $^1\varepsilon_\om$. And every  (conciliating) set that we can generate 
is of Wadge degree $<\,^1\varepsilon_\om$.  
In particular one cannot 
construct, from known $\om$-CFL of finite Borel rank, 
 an $\om$-context free language being a  ${\bf \Si^0_\om}$-complete Borel set. Indeed  
the Wadge degree of a  ${\bf \Si^0_\om}$-complete Borel set is $\,^1\varepsilon_{\om_1}$, 
i.e. the $(\om_1)^{th}$ fixed point of the operation of ordinal exponentiation 
of base $\om_1$, which is a much larger ordinal than $\,^1\varepsilon_\om$.  
 \nl However we are going to show in this section that there exist 
some ${\bf \Si^0_\om }$-{\bf complete}  $\om$-context free language, 
using other methods and results about sets of $\om^2$-words. 

\hs  The set  $\Si^{\om^2}$ is the set of $\om^2$-{\it words} over the finite alphabet $\Si$. 
It may also be viewed as the set of (infinite) $(\om \times \om)$-matrices 
whose coefficients are letters 
of $\Si$. If $x \in \Si^{\om^2}$ we shall write $x = (x(m, n))_{m\geq 1, n\geq 1}$.  
The infinite word $x(m, 1)x(m, 2)\ldots x(m, n)\ldots$ will be called 
the $m^{th}$ column of the $\om^2$-word $x$ and the infinite word 
$x(1, n)x(2, n)\ldots x(m, n)\ldots$ will be called 
the $n^{th}$ row of the $\om^2$-word $x$.  Thus an element of   $\Si^{\om^2}$ 
is completely determined by the (infinite) set of its columns or of its rows. 
\nl The set $\Si^{\om^2}$ is usually equipped with the product topology  of the discrete 
topology on $\Si$ (for which every subset of $\Si$ is an open set), see \cite{kec,pp01}.
For this topology on  $\Si^{\om^2}$, the basic open sets are the sets of $\om^2$-words 
with a fixed two-dimensional prefix. 
This topology may be defined 
by the following distance $d$. Let $x$ and $y$ be two  $\om^2$-words in $\Si^{\om^2}$ 
such that $x\neq y$, then  
$ d(x, y)=2^{-n},  ~~\mbox{   where  }$
$n=min\{p\geq 1 \mid \exists (i, j) ~x(i, j)\neq y(i, j) \mbox{ and } i+j=p\}$. 
\nl  Then the topological space $\Si^{\om^2}$ is homeomorphic to the above defined 
topological space  $\Si^{\om}$.  
The Borel hierarchy on  $\Si^{\om^2}$ is defined from open 
sets in the same manner as in the case of the topological space $\Si^\om$. The notion 
of ${\bf \Si^0_\alpha}$ (respectively ${\bf \Pi^0_\alpha}$)-complete sets are also 
defined in a similar way.  

 \hs   Recall now that the set 
$S = \{ x\in \{0, 1\}^{\om^2} \mid \exists m \exists^\infty n ~x(m, n)=1\}$, 
\noi where $\exists^\infty$ means "there exist infinitely many", 
is a ${\bf \Si^0_3}$-complete subset of $\{0, 1\}^{\om^2}$, \cite[p. 179]{kec}. 
It is the set of $\om^2$-words 
having at least one column in the ${\bf \Pi^0_2 }$-complete subset 
$\mathcal{R}=(0^\star.1)^\om$ of $\{0, 1\}^{\om}$. 
\nl In a similar manner we can prove the following result:

\begin{Lem}\label{lem1}
Let $L\subseteq \Sio$ be a ${\bf \Si^0_\om }$-subset of $\Sio$ which is of Borel rank $\om$. 
Then the set 
$\mathcal{L} = \{ x\in \Si^{\om^2} \mid \exists m ~~ x(m,1)x(m,2)\ldots x(m,n)\ldots \in L\}$
of $\om^2$-words over $\Si$ 
having at least one column in $L$ is a  ${\bf \Si^0_\om }$-complete subset of $\Si^{\om^2}$. 
\end{Lem}

\proo Let $L\subseteq \Sio$ be a ${\bf \Si^0_\om }$-subset of $\Sio$ of Borel rank $\om$
and let $\mathcal{L}_m$ 
be the  set of $\om^2$-words over $\Si$ 
having their $m^{th}$ column in $L$. 
 It is easy to check that for every integer $m\geq 1$ 
the set $\mathcal{L}_m$ is a  ${\bf \Si^0_\om }$-subset of $\Si^{\om^2}$. 
For that purpose, consider the function 
$i_m: \Si^{\om^2} \ra \Sio$ defined by $i_m(x)=x(m,1)x(m,2)\ldots x(m,n)\ldots $ for 
every $x\in \Si^{\om^2}$. Then it is easy to see that $i_m$ is a continuous function and that
$i_m^{-1}(L)=\mathcal{L}_m$ holds. Therefore  $\mathcal{L}_m$ is a 
 ${\bf \Si^0_\om }$-subset of $\Si^{\om^2}$ because the class ${\bf \Si^0_\om }$ is closed 
under inverse images  by  continuous functions. 
\nl   Thus the set 
$\mathcal{L}=\bigcup_{m\geq 1}\mathcal{L}_m$ of $\om^2$-words over $\Si$ 
having at least one column in $L$ is a countable union of ${\bf \Si^0_\om }$-sets so it is a 
${\bf \Si^0_\om }$-set because the class of ${\bf \Si^0_\om }$-subsets of $\Si^{\om^2}$ 
is closed under countable unions. 
\nl It remains to show that $\mathcal{L}$ is ${\bf \Si^0_\om }${\bf -complete}. Let then 
$S$ be a ${\bf \Si^0_\om }$-subset of $\Sio$. By definition of the class of 
${\bf \Si^0_\om }$-subsets of $\Sio$ there exist some subsets $A_i$, $i\in \mathbb{N}$, 
of $\Sio$ such that $S=\cup_{i\in \mathbb{N}}A_i$ and, for each integer $i$,~~
 $A_i \in {\bf \Pi^0_{j_i} }$ for some integer $j_i \geq 1$. But then 
it is well known that each set $A_i$ is the inverse image by some continuous function 
of the ${\bf \Si^0_\om }$-set $L$ which is of Borel rank $\om > j_i$: 
there exists a continuous function 
$f_i$ from $\Sio$ into $\Sio$ such that 
$f_i^{-1}(L)=A_i$  (this follows for example from the study of the Wadge hierarchy). 
\nl  Let  now $f$ be the  function  from $\Sio$ into $\Si^{\om^2}$ which is defined by 
$f(x)(m,n)=f_m(x)(n)$. The function $f$ is continuous because each function $f_i$ 
is continuous.
\nl    For $x\in \Sio$  $f(x)\in  \mathcal{L}$ iff the $\om^2$-word $f(x)$ has at least 
one column in the \ol~ $L$, i.e. iff there exists some integer $m \geq 1$ such that 
$$f_m(x)=f_m(x)(1)f_m(x)(2)\ldots f_m(x)(n) \ldots   \in L$$
iff  $\exists m \geq 1 ~~ x\in A_m$. Thus  $f(x)\in  \mathcal{L}$ iff 
$x\in S=\cup_{i\in \mathbb{N}}A_i$ so $S=f^{-1}(\mathcal{L})$. 
\nl  We have then proved that 
all ${\bf \Si^0_\om }$-subsets of $\Sio$ are inverse images by continuous functions of 
the  ${\bf \Si^0_\om }$-set $\mathcal{L}$ therefore $\mathcal{L}$ is 
a ${\bf \Si^0_\om }$-complete set.          \ep

 \hs  In order to simplify our  proofs we shall use in the sequel 
 the following variant of lemma \ref{lem1} which can be proved 
with a  slight  modification. 

\begin{Lem}\label{lem2}
Let $L\subseteq \Sio$ be a ${\bf \Si^0_\om }$-subset of $\Sio$ which is of Borel rank $\om$. 
Then the set 
$\mathcal{L}^{e} = 
\{ x\in \Si^{\om^2} \mid \exists m \geq 1 ~~ x(2m,1)x(2m,2)\ldots x(2m,n)\ldots \in L\}$
 of $\om^2$-words over $\Si$ 
having at least one column of {\bf even index } 
in $L$ is a  ${\bf \Si^0_\om }$-complete subset of $\Si^{\om^2}$. 
\end{Lem}

  \noi In order to use these results we shall firstly define a coding of $\om^2$-words 
over $\Si$ by $\om$-words over the alphabet  $(\Si\cup\{C, B\})$ where  
$C$ and $B$ are  new letters not in $\Si$.
  Let us call, 
for $x\in \Si^{\om^2}$ and $p$ an integer $\geq 2$:  
$$T^x_{p+1}=\{x(p,1), x(p-1, 2), \ldots , x(2, p-1), x(1,p)\}$$ 
\noi  the set of elements $x(m, n)$ with $m+n=p+1$ and 
$$U^x_{p+1}=x(p,1).x(p-1, 2) \ldots x(2, p-1).x(1,p)$$ 
the sequence formed by the concatenation of elements $x(m, n)$ of $T^x_{p+1}$ for 
increasing values of $n$. We also call 
$$V^x_{p+1}=(U^x_{p+1})^R=x(1,p).x(2, p-1)\ldots x(p-1, 2).x(p,1)$$ 
the reverse image of $U^x_{p+1}$. Thus $U^x_{p+1}$ and $V^x_{p+1}$ are 
finite non empty words over $\Si$ and $U^x_{2}=V^x_{2}=x(1,1)$.  

\hs   We shall code an $\om^2$-word $x\in \Si^{\om^2}$ by the $\om$-word $h(x)$ defined by
$$h(x)=V^x_2.C.U^x_3.B.V^x_4.C.U^x_5.B.V^x_6.C \ldots C.U^x_{2k+1}.B.V^x_{2k+2}.C\ldots $$
The word $h(x)$ begins with $x(1, 1)=V^x_{2}$ followed by a letter $C$; then the 
word $h(x)$  enumerates the elements of the sets $T^x_{p+1}$ for increasing values of the 
integer $p$. 
More precisely for every even integer $2k\geq 2$ the elements of $T^x_{2k+1}$ are 
enumerated by the sequence $U^x_{2k+1}$, followed by a letter $B$, followed by  
the elements  of $T^x_{2k+2}$, enumerated by the sequence $V^x_{2k+2}$, followed by a letter 
$C$, and so on \ldots

 \hs  Let then $h$ be the mapping from $\Si^{\om^2}$  into
 $(\Si\cup\{C, B\})^\om$ 
such that,  for every $\om^2$-word $x$ over the  alphabet $\Si$,  
$h(x)$ is the code  of the $\om^2$-word $x$ as defined above.
It is easy to see, from the definition of $h$ and of the  order of the enumeration 
of letters $x(m, n)$ (they are enumerated for  increasing values of $m+n$), 
that $h$ is a continuous function from $\Si^{\om^2}$  into 
$(\Si\cup\{C, B\})^\om$.

\begin{Lem}\label{lem3}
Let $\Si$ be a finite alphabet. If $\mathcal{L} \subseteq \Si^{\om^2}$ is  
${\bf \Si^0_\om }$-complete then 
$h(\mathcal{L}) \cup h(\Si^{\om^2})^-$
 is a ${\bf \Si^0_\om }$-complete subset of $(\Si \cup\{C, B\})^\om$.
\end{Lem}

\proo  The topological space $\Si^{\om^2}$  is compact 
thus its image by the continuous function 
$h$ is also a compact subset of the topological space 
$(\Si \cup\{C, B\})^\om$. 
The set  $h(\Si^{\om^2})$ is compact hence  it is a closed subset of 
$(\Si \cup\{C, B\})^\om$. Then its complement 
$(h(\Si^{\om^2}))^-$ 
  is an open (i.e. a ${\bf \Si^0_1}$) subset of $(\Si \cup\{C, B\})^\om$.
\nl 
On the other side the function $h$ is also injective 
thus it is a bijection from $\Si^{\om^2}$  onto 
$h(\Si^{\om^2})$. But a continuous bijection between two compact sets is an homeomorphism
therefore $h$ induces an homeomorphism between  $\Si^{\om^2}$  and  $h(\Si^{\om^2})$. 
By hypothesis $\mathcal{L}$ is a ${\bf \Si^0_\om}$-subset of $\Si^{\om^2}$  thus 
$h(\mathcal{L})$ is a 
${\bf \Si^0_\om}$-subset of  $h(\Si^{\om^2})$ (where Borel sets of the topological 
space $h(\Si^{\om^2})$ are defined from open sets as in the cases of the topological 
spaces $\Sio$ or $\Si^{\om^2}$). 
\nl The topological space $h(\Si^{\om^2})$ is a 
topological subspace of $ (\Si \cup\{C, B\})^\om$ and its 
topology  is induced by the topology on $(\Si \cup\{C, B\})^\om$: open sets 
of $h(\Si^{\om^2})$ are traces on $h(\Si^{\om^2})$ of open sets of 
$(\Si \cup\{C, B\})^\om$ and the same result holds for closed sets. Then 
one can easily show by induction that for every ordinal  $\alpha \geq 1$,  
${\bf \Pi^0_\alpha  }$-subsets 
(resp.  ${\bf \Si^0_\alpha  }$-subsets) of  $h(\Si^{\om^2})$ are traces on $h(\Si^{\om^2})$ of 
${\bf \Pi^0_\alpha  }$-subsets 
(resp.  ${\bf \Si^0_\alpha  }$-subsets) of  $(\Si \cup\{C, B\})^\om$, i.e. are 
intersections with $h(\Si^{\om^2})$ of 
${\bf \Pi^0_\alpha  }$-subsets 
(resp.  ${\bf \Si^0_\alpha  }$-subsets) of  $(\Si \cup\{C, B\})^\om$. 
\nl 
 But  $h(\mathcal{L})$ is a ${\bf \Si^0_\om}$-subset of  $h(\Si^{\om^2})$ hence there exists 
a  ${\bf \Si^0_\om}$-subset $T$ of  $(\Si \cup\{C, B\})^\om$ such that 
$h(\mathcal{L})=T \cap h(\Si^{\om^2})$. But $h(\Si^{\om^2})$ is a closed 
i.e. ${\bf \Pi^0_1}$-subset 
of $(\Si \cup\{C, B\})^\om$ and the class of ${\bf \Si^0_\om}$-subsets of 
$(\Si \cup\{C, B\})^\om$  is closed under finite intersection thus 
$h(\mathcal{L})$ is a ${\bf \Si^0_\om}$-subset of $(\Si \cup\{C, B\})^\om$.  
 \nl 
 Now  $h(\mathcal{L}) \cup  (h(\Si^{\om^2}))^-$ is the union of a ${\bf \Si^0_\om}$-subset 
and of a ${\bf \Si^0_1}$-subset of $(\Si \cup\{C, B\})^\om$ 
therefore it is a ${\bf \Si^0_\om}$-subset of $(\Si \cup\{C, B\})^\om$ 
because the class of 
${\bf \Si^0_\om}$-subsets of $(\Si \cup\{C, B\})^\om$ 
 is closed under finite (and even countable) union.  
\nl 
In order to prove that $h(\mathcal{L}) \cup  (h(\Si^{\om^2}))^-$ 
 is ${\bf \Si^0_\om}${\bf -complete} it suffices to remark 
that
$\mathcal{L}=h^{-1}[ h(\mathcal{L}) \cup  (h(\Si^{\om^2}))^- ]$
 This implies that $h(\mathcal{L}) \cup  (h(\Si^{\om^2}))^- $ is ${\bf \Si^0_\om}$-complete 
because  $\mathcal{L}$ is assumed to be ${\bf \Si^0_\om}$-complete.    \ep

\begin{Lem}\label{lem4}
Let $\Si$ be a finite alphabet and $h$ be the coding of $\om^2$-words over $\Si$ defined as 
above. Then $h(\Si^{\om^2})^-=(\Si\cup\{C, B\})^\om - h(\Si^{\om^2})$ is an $\om$-CFL. 
\end{Lem}

\proo   Remark first that $h(\Si^{\om^2})$  is the set of $\om$-words in 
$(\Si\cup\{C, B\})^\om $
which belong to 
$$\Si.C.\Si^2.B.\Si^3.C.\Si^4.B \ldots C.\Si^{2^k}.B.\Si^{2^{k+1}}.C \ldots $$

\noi In other words this is the set of words in $(\Si\cup\{C, B\})^\om $
 which are in $(\Sis.C.\Sis.B)^\om$ 
and have $k+1$ letters of $\Si$ between the $k^{th}$ and 
the $(k+1)^{th}$ occurrences of letters in $\{C, B\}$. It is now easy to see 
that  the complement of the set $h(\Si^{\om^2})$ of codes of 
$\om^2$-words over $\Si$ is the union of the sets $\mathcal{C}_1$ and $\mathcal{C}_2$ where:

\begin{itemize} 
\ite $\mathcal{C}_1 = (\Si\cup\{C, B\})^\om - (\Sis.C.\Sis.B)^\om$ hence $\mathcal{C}_1$ 
is the complement  of the \orl~ $(\Sis.C.\Sis.B)^\om$ so it is also an \orl~  thus 
$\mathcal{C}_1$ is  an $\om$-CFL. 
\ite $\mathcal{C}_2$ is the set of $\om$-words over $(\Si\cup\{C, B\})$ which may 
be written in the form $w.u.C.t$ or $w.u.B.t$ where $w\in (\Sis.\{C,B\})^k$, for $k\geq 0$, 
 and $u\in \Sis$ and 
$|u|\neq k+1$ and $t\in (\Si\cup\{C, B\})^\om$. It is easy to show that 
$$\mathcal{C}=\{w.u \mid  w\in (\Sis.\{C,B\})^k \mbox{ for an integer } k\geq 0 \mbox{ and }
 u\in \Sis \mbox{ and } |u|\neq k+1 \}$$
\noi is a context free finitary language, thus 
$\mathcal{C}_2=\mathcal{C}.\{C, B\}.(\Si\cup\{C, B\})^\om$ is an $\om$-CFL.
\end{itemize} 

\noi Now $h(\Si^{\om^2})^- = \mathcal{C}_1 \cup \mathcal{C}_2$ is an $\om$-CFL because  
 $CFL_{\om}$ is closed under finite union.  \ep 

 \hs  Let $L\subseteq \Sio$ be an $\om$-CFL  over the alphabet $\Si$ and 
$$\mathcal{L}^{e}=\{ x\in \Si^{\om^2} \mid \exists m \geq 1 
~~ x(2m,1)x(2m,2)\ldots x(2m,n)\ldots \in L\}$$  
We cannot show that $h(\mathcal{L}^{e})$ is an $\om$-CFL but we shall find  an 
$\om$-CFL $\mathcal{C}^{e} \subseteq (\Si \cup\{C, B\})^\om$ such that,  
 for every  $\om^2$-word $x \in \Si^{\om^2}$, 
$h(x) \in \mathcal{C}^{e}$ if and only if  $x \in \mathcal{L}^{e}$. 
We are now going to describe the \ol~ $\mathcal{C}^{e}$. 
  A word $y  \in (\Si \cup\{C, B\})^\om$ is in $\mathcal{C}^{e}$  
if and only if it is in the form
$$y=U_k.t(1).u_1.B.v_1.t(2).w_1.C.z_1.t(3) \ldots$$
$$\ldots t(2n+1).u_{n+1}.B.v_{n+1}.t(2n+2).w_{n+1}.C.z_{n+1}.t(2n+3) \ldots$$
where $k$ is an integer $\geq 1$, ~~~$U_k  \in (\Sis.C.\Sis.B)^{k-1}.(\Sis.C)$, 
and for all integers $i\geq 1$, $t(i)\in \Si$ and  $u_i, v_i, w_i, z_i \in \Sis$ and 
$$|v_i|=|u_i|  ~~~~ \mbox{ and  }  ~~~~|z_i|=|w_i|+1$$ 
and the $\om$-word $t=t(1)t(2)\ldots t(n)\ldots $ is in the \ol~ $L$.

\hs   We now state the following result. 

\begin{Lem}\label{lem5}
Let $L\subseteq \Sio$  and let 
$\mathcal{L}^{e}  \subseteq \Si^{\om^2}$ and  
$\mathcal{C}^{e} \subseteq (\Si \cup\{C, B\})^\om$  be defined as above. Then 
$ \mathcal{L}^{e}= h^{-1} (\mathcal{C}^{e})$, i.e. ~~
$\fa x \in \Si^{\om^2} ~~~~ h(x) \in \mathcal{C}^{e} 
\longleftrightarrow x\in \mathcal{L}^{e}$.
\end{Lem}

\proo Let $L\subseteq \Sio$ be an omega language over the alphabet $\Si$ and let 
$\mathcal{L}^{e}  \subseteq \Si^{\om^2}$ and  
$\mathcal{C}^{e} \subseteq (\Si \cup\{C, B\})^\om$  be defined as above. 
Assume now that an $y\in \mathcal{C}^{e}$, written in the above form, is the code $h(x)$ of  
an $\om^2$-word $x \in \Si^{\om^2}$, then 
$t(1).u_1 = U_{2k+1}^x = x(2k,1)x(2k-1,2)\ldots x(1,2k)$. 
So in particular 
$x(2k,1)=t(1)$
holds.   Next 
$v_1.t(2).w_1 = V_{2k+2}^x$
then 
$x(2k,2)=t(2)$
 holds because the elements of $T^x_{2k+2}$ and the 
elements of $T^x_{2k+1}$ are enumerated in reverse orders 
in the code of $x$ and because $|u_1|=|v_1|$. 
 Then $|z_1|=|w_1|+1$ implies that 
$x(2k,3)=t(3)$.  
 \nl  By construction this phenomenon will happen  further. One can easily show by 
induction on integers $n$ that letters $t(n)$ are successive letters 
of the $(2k)^{th}$ column of $x$.  For that purpose assume that for some integer 
$n\geq 1$ it holds that $t(2n+1)=x(2k, 2n+1)$. By definition of the code $h(x)$ 
we know that 
$$U_{2k+2n+1}^x = x(2k+2n,1)x(2k+2n-1,2)\ldots x(1,2k+2n)=z_n.t(2n+1).u_{n+1}$$
\noi and  $t(2n+1)=x(2k, 2n+1)$ implies that $|u_{n+1}|=(2k+2n)-(2n+1)=2k-1$. 
Thus $|v_{n+1}|=|u_{n+1}|=2k-1$. But it holds also that 
$$V_{2k+2n+2}^x = x(1,2k+2n+1)x(2,2k+2n)\ldots x(2k+2n+1,1)=v_{n+1}.t(2n+2).w_{n+1}$$
\noi therefore $t(2n+2)= x(2k, 2n+2)$ and $|w_{n+1}|=(2k+2n+1)-(2k)=2n+1$. 
But $|z_{n+1}|=|w_{n+1}|+1=2n+2$ and 
$$U_{2k+2n+3}^x = x(2k+2n+2,1)x(2k+2n+1,2)\ldots x(1,2k+2n+2)=z_{n+1}.t(2n+3).u_{n+2}$$
\noi thus $t(2n+3)=x(2k, 2n+3)$. 

\hs Then we have proved by induction that 
$$t=t(1)t(2)\ldots t(n)\ldots = x(2k,1)x(2k,2)\ldots x(2k,n)\ldots$$ 

 \noi Thus if a code $h(x)$ of  an $\om^2$-word $x \in \Si^{\om^2}$ is in $\mathcal{C}^{e}$ then 
$x$ has a column of even indice in $L$, i.e. $x\in \mathcal{L}^{e}$. 
Conversely it is easy to see that every code $h(x)$ of $x\in \mathcal{L}^{e}$ may be written 
in the above form of a word in  $\mathcal{C}^{e}$ (remark that if $x\in \mathcal{L}^{e}$ 
has several columns of even indices in $L$ 
then $h(x)$ has several written forms as above, the integer 
$k$ determining one of the columns of even index of $x$ being in $L$).  
\nl   Then we have proved that for every  $\om^2$-word $x \in \Si^{\om^2}$
$h(x) \in \mathcal{C}^{e}$ if and only if  $x \in \mathcal{L}^{e}$. 
\ep 

\begin{Lem}\label{lem6}
Let $L\subseteq \Sio$ be an $\om$-CFL over the alphabet $\Si$ and let 
$\mathcal{L}^{e}  \subseteq \Si^{\om^2}$ and  
$\mathcal{C}^{e} \subseteq (\Si \cup\{C, B\})^\om$  be defined from $L$ as above. Then 
$\mathcal{C}^{e}$ is an $\om$-CFL over the alphabet $\Si \cup\{C, B\}$. 
\end{Lem}

\proo Assume that $L\subseteq \Sio$ is an $\om$-CFL and let 
$\mathcal{L}^{e}  \subseteq \Si^{\om^2}$ and  
$\mathcal{C}^{e} \subseteq (\Si \cup\{C, B\})^\om$  be defined from $L$ as above. 
Let $\mathcal{D}_1$ and  $\mathcal{D}_2$ be the following finitary languages:
$$\mathcal{D}_1 = \{ u.B.v \mid u, v \in  \Sis \mbox{ and } |u|=|v| \},$$
$$\mathcal{D}_2 = \{ w.C.z \mid w, z \in  \Sis \mbox{ and } |z|=|w|+1 \}.$$
It is easy to see that $\mathcal{D}_1$ and  $\mathcal{D}_2$ are context free 
finitary languages over the alphabet $\Si \cup \{C, B\}$ thus 
$\mathcal{D} = \mathcal{D}_1 \cup \mathcal{D}_2$ is also a context free 
finitary language. 

 \hs  Recall now the definition of substitution in languages:
 a substitution $f$ is defined by a mapping 
$\Si\ra P(\Ga^\star)$, where $\Si =\{a_1, a_2 \ldots ,a_n\}$  
and $\Ga$ are two finite alphabets, 
$f: a_i \ra L_i$ where $\fa i\in [1;n]$, $L_i$ is a finitary language over the alphabet $\Ga$.
This mapping is extended in the usual manner to finite words:
$f(x(1) \ldots x(n))= \{u_1 \ldots u_n \mid  \fa i\in [1;n]~u_i\in f(x(i))\}$,  
 where $x(1)$, \ldots , $x(n)$ are letters in $\Si$, 
and to finitary languages $E\subseteq \Sis$: 
$f(E)=\cup_{x\in E} f(x)$. 
\nl The substitution $f$ is called $\lambda$-free if for every $i\in [1;n]$  $L_i$ does not 
contain the empty word. In that case the mapping $f$ may be extended to $\om$-words:
$f(x(1)\ldots x(n)\ldots )= \{u_1\ldots u_n \ldots  \mid  \fa i \geq 1 ~u_i\in f(x(i))
\}$;  
and to \ol s $E\subseteq \Sio$ by  $f(E)=\cup_{x\in E} f(x)$. 
\noi  Let $\mathbb{F}$ be a family of languages, if for every $i\in [1;n]$ the language 
$L_i$ belongs to  $\mathbb{F}$ 
the substitution $f$ is called a $\mathbb{F}$-substitution.

 \hs   Let then $g$ be the substitution $\Si\ra P((\Si \cup \{C, B\})^\star)$ defined by:
$a \ra a.\mathcal{D}$ where $\mathcal{D}$ is the context free language defined above. 
Then $g$ is a $\lambda$-free substitution. But the languages $a.\mathcal{D}$ are context free 
and $CFL_\om$ is closed under $\lambda$-free context free substitution \cite{cg} thus the 
\ol~ $g(L)$ is context free. 

\hs   The class $CFL_\om$ is closed under intersection with \orl s, therefore the \ol~ 
$g(L) \cap (\Sis.B.\Sis.C)^\om$
is context free. But it is  easy to see that 
$$\mathcal{C}^{e} = (\Sis.C.\Sis.B)^\star.(\Sis.C).[ g(L) \cap (\Sis.B.\Sis.C)^\om ].$$ 
\noi Thus $\mathcal{C}^{e}$ is an $\om$-context free language 
because the class $CFL_\om$ is closed under 
left concatenation by regular (finitary) languages.    
  \ep

\hs   We can now state the main result of this section.

\begin{The}\label{sio}
There exist some  $\om$-context free languages which are ${\bf \Si^0_\om }$-complete 
Borel sets. 
\end{The}

\proo We already know that there exist some  $\om$-context free languages which are 
${\bf \Si^0_\om }$-sets of Borel rank $\om$. 
   Let then $L$ be such an $\om$-CFL and let  
$\mathcal{L}^{e} \subseteq \Si^{\om^2}$ 
and $\mathcal{C}^{e}\subseteq (\Si \cup\{C, B\})^\om$ defined from $L$ as above.  
 By Lemma \ref{lem2} the set $\mathcal{L}^{e}$ is a ${\bf \Si^0_\om }$-complete subset 
of $\Si^{\om^2}$ then by Lemma \ref{lem3} the \ol~ 
$h(\mathcal{L}^{e}) \cup  h(\Si^{\om^2})^-$
is a ${\bf \Si^0_\om }$-complete subset of $(\Si \cup\{C, B\})^\om$.  
   But Lemma \ref{lem5} states that $ \mathcal{L}^{e}= h^{-1} (\mathcal{C}^{e})$, 
and this implies that
$h(\mathcal{L}^{e}) \cup  h(\Si^{\om^2})^- = \mathcal{C}^{e} \cup  h(\Si^{\om^2})^- $. 
 On the other side we have proved in Lemmas \ref{lem4} and \ref{lem6} that the \ol s 
$h(\Si^{\om^2})^- $ and $\mathcal{C}^{e}$ are context free. Thus their union is an 
$\om$-context free language which is a 
${\bf \Si^0_\om }$-complete subset  of $(\Si \cup\{C, B\})^\om$.  
\ep 

\hs 
We know that the Wadge degree of the ${\bf \Si^0_\om }$-complete Borel 
set $\mathcal{C}^{e} \cup  h(\Si^{\om^2})^- $ 
is the ordinal  $\,^1\varepsilon_{\om_1}$. 
On the other hand it is easy to see that this set has same degree if we consider it as 
a conciliating set. 
So we can now state the following result, improving Theorem \ref{thewa} of preceding section. 

\begin{The}
The length of the conciliating hierarchy of infinitary context free languages,  
 which are Borel of rank $\om$, is strictly greater  than  $\varepsilon_\om$. 
 The length of the Wadge hierarchy of $\om$-context free languages  in 
$CFL_{\om} \cap {\bf \Si^0_\om }$ 
 is strictly  greater  than  $\varepsilon_\om $. 
\end{The}

\section{Concluding remarks}

\noi We have improved previous results on Wadge and Borel hierarchies of 
$\om$-context free languages. We have  proved the existence of 
$\varepsilon_\om$ Wadge degrees of $\om$-context free languages.  
And we have also given an inductive construction of an $\om$-context free language 
in each of these degrees and also of a B\"uchi or Muller pushdown automaton accepting it, 
using the previous work of Duparc on the Wadge hierarchy of Borel sets. 
A challenging question is to determine all the Borel ranks and the Wadge degrees of 
$\om$-context free languages.  

\section*{Acknowledgements} 
Thanks to Jean-Pierre Ressayre and  Jacques Duparc 
for many useful discussions about Wadge and 
Wagner Hierarchies and to the anonymous referees for useful comments 
on a preliminary version of this paper. 

\begin{footnotesize}

\end{footnotesize}

\end{document}